\documentclass[aps,twocolumn,showpacs,superscriptaddress,nofootinbib]{revtex4-1}
\pdfoutput=1
\usepackage{dcolumn}   
\usepackage{bm}        
\usepackage{amssymb}   
\usepackage[none]{hyphenat}

\usepackage[usenames,dvipsnames]{xcolor}

\usepackage[utf8]{inputenc}
\usepackage{amsmath,amsfonts,amssymb,amscd,amsthm,xspace}
\usepackage{slashed}
\usepackage{epsfig}
\usepackage{epstopdf}
\usepackage{mathrsfs}

\usepackage{dsfont}
\usepackage{graphicx}
\usepackage{color}
\usepackage{hyperref}
\hypersetup{urlcolor=black, colorlinks=false}

\newcommand\beq{\begin{equation}}
\newcommand\eeq{\end{equation}}
\newcommand\beqa{\begin{eqnarray}}
\newcommand\eeqa{\end{eqnarray}}
\newcommand{\dd}{\mathrm{d}}
\newcommand{\nn}{\nonumber}

\newcommand\beqal{\begin{align}}
\newcommand\eeqal{\end{align}}

\newcommand{\la}{\langle}
\newcommand{\ra}{\rangle}

\newcommand{\Psip}{\Psi^{(+)}}
\newcommand{\Psim}{\Psi^{(-)}}
\newcommand{\Psipm}{\Psi^{(\pm)}}
\newcommand{\psip}{\psi^{(+)}}
\newcommand{\psim}{\psi^{(-)}}
\newcommand{\tpsip}{\psi'^{(+)}}
\newcommand{\tpsim}{\psi'^{(-)}}

\newcommand{\bv}[1]{\mathbf{#1}}

\begin{document}



\title{Localization for Dirac fermions}
\author{Aleksander M. Kubicki}
\affiliation{Instituto de F\'isica Fundamental, CSIC, Serrano 113-B, 28006 Madrid, Spain}
\affiliation{Dep. Análisis Matemático, Fac. Matemáticas, Universidad de Valencia, Dr. Moliner 50, 46100 Burjassot, Spain}
\author{Hans Westman}
\affiliation{Instituto de F\'isica Fundamental, CSIC, Serrano 113-B, 28006 Madrid, Spain}
\affiliation{Alef Omega, LLC
1035 Pearl Street,
80302 Boulder CO, USA}
\author{Juan Le\'on}
\email{juan.leon@csic.es}\affiliation{Instituto de F\'isica Fundamental, CSIC, Serrano 113-B, 28006 Madrid, Spain}

\date{\today}
\begin{abstract}
This work is devoted to incorporating into QFT the notion that particles and hence the particle states should be localizable in space. It focuses on the case of the Dirac field in 1+1 dimensional flat spacetime, generalizing  a recently developed formalism for scalar fields. This is achieved exploiting again the non-uniqueness of quantization process. Instead of elementary excitations carrying definite amounts of energy and momentum, we construct the elementary excitations of the field localized (at some instant) in a definite region of space. This construction not only leads to a natural notion of localized quanta, but also provides a local algebra of operators. Once constructed, the new representation is confronted to the standard global (Fock) construction. In spite of being unitarily inequivalent representations, the localized operators are well defined in the conventional Fock space. By using them we dig up the issues of localization in QFT showing how the globality of the vacuum state is responsible for the lack of a ``common sense localization'' notion for particles in the standard Fock representation of QFT and propose a method to meet its requirements.

\end{abstract}
\maketitle

\section{Introduction}
\label{Intro}

There is a basic obstacle to incorporate into QFT the notion that particles and  particle states can be localized in space. The problem appears when trying to define states which may be strictly localized \footnote{  In QFT, a state $|\Psi_V\ra$ is \emph{strictly localized} within the region $V\subset \mathds{R}^3$ 
 if the expectation value of any local operator $\mathcal{Q}(x)$ at a point $x$ outside the region (i.e. x $\not\in V$) is identical to that in the vacuum state \cite{Knight,Licht1963}:
\beq
\la \Psi_{V} | \mathcal{Q}(x)  | \Psi_V\ra= \la 0 |  \mathcal{Q}(x)  | 0\ra \text{  if  }\ x\ \not\in V. \nn
\eeq There exists several other  localization notions among the literature accounting for states which \emph{differ only locally} from the vacuum, but all of them are proved to be equivalent to the \emph{strict localization} introduced above \cite{Wallace2006,Deutsch2000}.   } within $\mathcal{O}$, a region $V$ of space during the time interval $(t_1,t_2)$.
 The idea is that  by applying the operators of the local algebra $\mathcal{R}(\mathcal{O})$ to the vacuum state $|0\rangle$ one would generate a subspace $\mathcal{H}_{V}$ of local states. However, this does not work because $\mathcal{H}_{V}$ turns out to be the whole Hilbert space \cite{HaagSwieca1965,ReedSchlieder1961}. Simple  questions as ``how many particles are there?'' when referred to some finite region of space, or ``how are they spatially distributed?'' can hardly be confronted in this background if not by indirect means. Nonetheless, dealing with asymptotic states saves scattering theory from these problems. On the other hand, QFT evades these problems, at least partially, smearing the fields with compact support functions, something that in any case would be necessary to avoid ill-defined expressions. Hence, there is a strong  temptation to overlook these issues of localizability as harmless. But beware, the very notion of particle depends of them \cite{BialynickiBirula1998,Terno2014}. A strong case of this, by no means unique, is the Unruh effect \cite{Fulling1973}, where a (thermal) bath of particles appears due the accelerated observer's  inability to access to the part of the (vacuum) state beyond the horizon. The problem naturally manifests in the context of curved spacetimes \cite{Wald} and beyond that \cite{RovelliBook,ThiemannBook}, in Quantum Gravity, where horizons will act on operators as continental divides.  In short, the notion of particle is not objective in QFT;  in particular, it depends of its space localization.

The early attempts to overcome the difficulties alluded to above consisted in the recipe of not applying the whole local algebra of operators $\mathcal{R}(\mathcal{O})$ on the vacuum state but, instead, applying only the unitary (or isometric) operators of this algebra to obtain localized states \cite{Knight,Licht1963}. However, not only mathematical, but also physical difficulties persisted. It was independently shown that in quantum theory, if there is a lower bound on the system's energy, the wavefunction of any state with localized particles spreads superluminally violating Lorentz causality \cite{Hegerfeldt1985,Hegerfeldt1998}. Besides, there is no relativistic quantum theory with particles that can be sharply or unsharply localized in finite spatial regions \cite{Malament,HalvorsonTh,IwoSharp}. Beyond that, in contrast to non relativistic theories, relativistic quantum field theories do not admit systems of local number operators \cite{HalvorsonTh}. Hence, not only states, but even operators collude against the notion of localization. In short, the idea that QFT describes particles seems at odds with its difficulty to deal with the localizability of these entities.

A suitable scheme has been recently proposed for the description of localized quanta \cite{localquanta} that evolves from some ideas presented in  \cite{ColosiRovelli2004}. This approach focuses on the creation and annihilation of elementary excitations localized in finite regions of space, i.e., associated to modes vanishing outside these regions, instead of elementary excitations with sharp momentum. This provides a representation of the quantum field complementary to the standard Fock construction, resembling the complementarity between the position and momentum representations of nonrelativistic quantum mechanics. However both representations, that of \cite{localquanta} and that of Fock, fail to be unitarily equivalent. This fact, a common feature in field theories, comes from the infinite number of degrees of freedom inherent to the fields, which prevents from applying the Stone-Von Neumann theorem \cite{Stone1930,v.Neumann1931}. Within this localizing approach there is room for localized states, local number operators,  and a variety of tools to dig up the issues of localization. It clearly spells out how it is the global nature of the quantum vacuum the responsible for the difficulties with the ``common sense localization'' notion for particles in the standard Fock representation of QFT. In poor words (and simple notation), the root of the problem resides in that, while it is possible to decompose the vacuum as a product state such as $|0\rangle = \bigotimes_k |0_k\rangle$, there is no way for $|0\rangle = \bigotimes_V |0_V\rangle$, with $k$ and $V$ labelling momenta and space regions respectively.

 The construction developed in \cite{localquanta} was endowed in \cite{whatdoesitmeans}  with an operational interpretation. It was the tool used there to analyze the creation of quanta from the vacuum when slamming a mirror in a cavity and its relation with the dynamical Casimir effect. It has also been used in \cite{Dragan2016} to investigate the entanglement of nonvacuum Gaussian states and how the entanglement between spatial areas and the entanglement in the global basis of plane waves relate to each other.

In this work we extend the construction of \cite{localquanta} to fermion fields. Specifically, we focus here on the case of Dirac fermions in one space dimension. As in the boson case, we consider quanta whose wave function at some instant of time -the initial time- vanish outside a finite region of space, i.e. out of a finite segment of the real line. The complete set of fermion modes necessary to describe one of these, otherwise arbitrary, initial conditions are provided by the  modes whose vector current vanishes at both ends of the segment and beyond. This customizes what was done in \cite{localquanta}. We followed here the treatment of Boundary Conditions of the MIT bag model \cite{MITbag1} that has been used for the fermion Casimir effect \cite{Bordag20011} and also in some questions of Relativistic Quantum Information \cite{Friis}. Besides, this also simplifies further generalizations to 3 space dimensions.
Statistics are fully accounted for in our treatment by the use of canonical anticommutation relations (CARs), this produces subtle effects even in this case of unitarily inequivalence. It is shown that the fermionic statistics of the system manifest by constraining the contribution of high frequency modes to the mean values and correlations of the local operators we construct. As a consequence, the number of \emph{local} excitations with fixed quantum numbers in any Fock state remains bounded by 1 as dictated by the Pauli Exclusion Principle.

The present text is structured as follows: Sec.\ref{sec1}  contains the basics of the 1+1-dim. Dirac field confined within a cavity, with special emphasis on the peculiarities brought about by the boundary conditions at both ends. Sec.\ref{sec2} deals with the modes initially localized in a part of the cavity and their  quantization; here we introduce the localized excitations of the fermion field. Next, in Sec.\ref{sec3} we discuss the relation between both kinds - localized and global - of fermion quanta. Then, we introduce the Bogoliubov transformation between them and show that both representations are unitarily inequivalent. Sec.\ref{sec4} is devoted to the application of the formalism presented here to the study of local features in the vacuum state of the ordinary QFT. Finally, we conclude summing up the work we present here and spotting future directions for further development. Technical details that prevent a fluid reading of the main text have been put aside to the appendices \ref{modes}, \ref{BogAp}, \ref{LimAp}.

\section{1+1-D Dirac field in a cavity}
\label{sec1}

\subsection{The 1+1-D Dirac field}

Consider $\Psi(\bv{x}, t)$ being a Dirac spinorial field, i.e., an object in the so-called representation $(\frac{1}{2},0)\oplus(0,\frac{1}{2})$ of the Lorentz group satisfying the Dirac equation:
\beq
\left( i \slashed{\partial} - m \right) \Psi(\bv{x}, t) = 0,
\eeq
being $m$ the mass associated to the field. The Feynmann's slash notation was adopted to denote contraction with the gamma matrices.

In this essay natural units $\hbar=c=k_B=1$ and metric signature $g_{00} = 1$ will be used.

The probability current associated to the field $\Psi$ is:
\beq
j^{\mu}(\bv{x}, t) = \overline{\Psi}(\bv{x}, t) \gamma^{\mu} \Psi(\bv{x}, t),
\eeq
whose first component  induces the definition of the inner product in the space of classical solutions, $\mathcal{S}$:
\beq
(\Psi|\Psi') = \int \dd^3 x \ \overline{\Psi}(\bv{x}, t) \gamma^0 \Psi'(\bv{x}, t) = \int \dd^3 x \ \Psi^{\dagger}(\bv{x}, t) \Psi'(\bv{x}, t). \label{innerproduct}
\eeq

This inner product is conserved through the (unitary) evolution generated by the Dirac Hamiltonian:
\beq
H_D = \left( -i \alpha^j \frac{\dd}{\dd x^j} + m \beta \right), \quad \beta= \gamma^0,\quad \alpha^j= \beta \gamma^j,\label{HD1}
\eeq
and makes it self-adjoint with respect to that inner product. Then, the spectrum of this Hamiltonian is composed by orthogonal eigenspinors which constitute a complete set in $\mathcal{S}$ in virtue of the spectral theorem.

For simplicity, we restrict to the 1+1 dimensional problem. In this case, properly speaking there is no room for rotations and therefore no notion of spin, but still two spinor representations associated to the boosts of the Lorentz group $\mathrm{O}(1,1)$ \cite{Gitman}. The standard four component spinorial representation in $3+1$ dimensions reduces to a two component spinor, which in the Dirac representation reads:
\beq
\Psi =\frac{1}{\sqrt{2}}\left(
\begin{array}{c}
 \psi _L+\psi _R \\
 \psi _L-\psi _R
\end{array}
\right)\equiv \left(
\begin{array}{c}
 \phi  \\
 \chi
\end{array}
\right),
\eeq
choosing consistently the following representation of the  $\gamma$ matrices:
\beq
\gamma ^0=\left(
\begin{array}{cc}
 1 & 0 \\
 0 & -1
\end{array}
\right),\text{  }\gamma ^1=\left(
\begin{array}{cc}
 0 & 1 \\
 -1 & 0
\end{array}
\right).
\eeq

Above, $\psi_L,\  \psi_R$ are the left-handed and right-handed components in the Weyl representation.

Setting this representation, independent plane wave solutions to the Dirac equation can be written as:
\beqa
U_p(x,t) &=& e^{-i ( \omega t - p x)} u(p),   \label{pwmas}\\
V_p(x,t) &=& e^{i ( \omega t - p x)} v(p) ,  \label{pwmenos}
\eeqa
where the spinors
\beq
u(p)=\sqrt{\frac{\omega +m}{2\omega }}\left(
\begin{array}{c}
 1 \\
 \frac{p}{\omega +m}
\end{array}
\right)\text{  },\text{  }v(p)=\sqrt{\frac{\omega +m}{2\omega }}\left(
\begin{array}{c}
 \frac{p}{\omega +m} \\
 1
\end{array}
\right),\nn \label{eigenspinors}
\eeq
make \eqref{pwmas}, \eqref{pwmenos} eigenstates of the Hamiltonian \eqref{HD1} with eigenvalue $\omega$, $-\omega$ respectively (being $\omega = \sqrt{p^2 +m^2}$). They are normalized according to the relations:
\begin{gather}
u^{\dagger}(p)u(p) = 1,\qquad  v^{\dagger}(p)v(p) = 1,\nn\\
 u^{\dagger}(p)v(-p)=0.\label{uvprods}
\end{gather}

Then, any solution to the Dirac equation, $\Psi\in\mathcal{S}$, can be spanned in terms of \eqref{pwmas} and \eqref{pwmenos}:
\beq
\Psi = \int_{-\infty}^{\infty}  \frac{\dd p}{\sqrt{2 \pi}}  \ \left( a(p) e^{-i ( \omega t - p x)} u(p) + b^*(p) e^{i ( \omega t - p x)} v(p) \right),\nn
\eeq
where $a(p), \: b^*(p)$ are complex functions playing the role of expansion coefficients.

To close this preliminar section, let us point out that the notation introduced in \eqref{innerproduct} for the inner product encourages to denote vectors in the Hilbert space $\mathcal{S}$ as $|\Psi)$. In the same way, elements $(\Psi|$ can be seen as forms in this space.

\subsection{Stationary solutions in a cavity}

Now we turn into finding the eigenspinors of the Dirac Hamiltonian for a Dirac field constrained to a cavity, in our case, a segment of the real axis, $\mathcal{I}=(0,R)$.

The field is constrained within the cavity by means of boundary conditions (BCs, from now on), as usual. The naiv\"{e} choice of Dirichlet conditions at the boundary of $\mathcal{I}$:
\beq
\Psi(x=0,t) = 0 = \Psi (x=R,t),
\eeq
leads uniquely to the trivial solution $\Psi(x,t) = 0$\footnote{It is easy to understand this situation. The Dirac equation (in  one-dimensional case) is composed by  two coupled first order equations for two fields (the two components of $\Psi$), and imposing  the vanishing values of both fields and their stationarity is too restrictive.}. Thus, this situation lacks of any physical interest and we have to consider more general localization schemes.

Then, the physical criterion to describe a Dirac field in a cavity will consist in imposing the vanishing value of the \emph{probability} current through the boundaries:
\beq
j^1(x=0,t) = 0 = j^1(x=R,t).\label{BC}
\eeq

This is achieved imposing at the boundary, $\partial \mathcal{I}$, the following condition:
\beq
\left(1+i \slashed{\bv{n}} \right) \Psi |_{\partial} =0,\label{BCMIT}
\eeq
 being $\bv{n}$ the outward normal vector to the boundary. These BCs respect the self-adjointness  of the Hamiltonian and the CPT symmetry of the field\footnote{ We understand here the parity transformation as the symmetry respect to the center of the cavity.} \cite{ven2}.

In our case, \eqref{BCMIT} means:
\beq
\Psi|_{x=0} = i \gamma^1 \Psi|_{x=0}, \qquad  \Psi|_{x=R} = - i \gamma^1 \Psi|_{x=R}.
\eeq

The conditions \eqref{BCMIT} are also the BCs derived from the M.I.T. bag model, which originally was developed as a hadronic model in the 1970's \cite{MITbag1}. These BCs can be seen as the result of coupling the free field in the cavity with an infinitely massive field outside. In that sense, they are analogue to the Dirichlet BCs in non-relativistic quantum mechanics.
 Furthermore, this model is widely used to describe  Dirac fields in  finite regions in the study of Casimir effects \cite{Bordag20011} and recently in relativistic quantum information  \cite{Friis}.  In  3 dimensional situations  with spherical symmetry, the radial problem is separated from the angular momentum sector. Then, the extension of the 1 dimensional treatment presented here to the more realistic 3 dimensional problem simply follows from adding the angular part of the problem (including the spin) and adapting the treatment of the spatial part given by us to the radial component. Our treatment also seems suitable to further generalizations dealing with higher spin fields, where a unified treatment of bag like BC were recently developed, \cite{StokesBennett2}, and applied to the study of the Casimir effect, \cite{StokesBennett1}.

Now, we construct the orthonormal stationary modes in our cavity, i.e., we solve the eigenproblem for the Dirac hamiltonian \eqref{HD1} in the domain:
\beq
 \left\lbrace \Psi \in \mathcal{S} \ \ | \ \ \left(1+i \slashed{\bv{n}}\right) \Psi |_{\partial \mathcal{I}} =0 \right\rbrace\equiv\mathcal{S}_{\mathcal{I}}.
\eeq

The detailed computation of this spectrum is deferred to the appendix \ref{modes}.

After normalizing the eigenmodes obtained with respect to the inner product \eqref{innerproduct} we finally can write:
\begin{widetext}\vspace{-4.5mm}\beqa
\Psip_I (x,t)&=& \sqrt{\frac{\Omega_I^2}{2 R (\Omega_I^2 + {m}/{R})}}   e^{-i \Omega_I t}     \left(e^{i (P_I x + \Delta_I)}\: u(P_I) - e^{-i (P_I x+\Delta_I)}\: u(-P_I) \right),\label{GM+}\\
\Psim_I(x,t) &=&\sqrt{\frac{\Omega_I^2}{2 R (\Omega_I^2 + {m}/{R})}}   e^{i \Omega_I t}    \  \left(e^{-i (P_I x + \Delta_I)}\: v(P_I) - e^{i (P_I x+\Delta_I)}\: v(-P_I) \right), \label{GM-}
\eeqa \vspace{-4.5mm} \end{widetext}
where:
\beq
\Omega_I= \sqrt{P_I^2 + m^2},  \qquad \Delta_I = \arctan\left(\frac{P_I}{\Omega_I + m}\right).\nn
\eeq

Here, the discrete (but infinite) spectrum $\lbrace P_I \rbrace$ is determined by the solutions to the trascendental equation:
\beq
\tan(P_I R) = - \frac{P_I}{m}.\label{Peq}
\eeq

From now on we keep capital letters for momenta, frequencies and labels corresponding to this set of modes.

These modes fulfil the  orthonormality relations:
\begin{gather}
(\Psip_I | \Psip_J ) = \delta_{I J}, \quad (\Psim_I | \Psim_{J} ) = \delta_{I J}\nn\\
(\Psip_I | \Psim_J ) = 0,                                                              \label{orthonormality}
\end{gather}

and decompose the identity as follows:
\beq
\mathds{1} = \sum_I \left( |\Psip_I)(\Psip_I| + |\Psim_I)(\Psim_I| \right).\label{leftid}
\eeq
 Hence, they built up an orthonormal basis\footnote{ As we have just noted before, the orthogonality is guaranteed by the self-adjointness of $H_D$ and the completeness follows from the spectral theorem.}  of $\mathcal{S}_{\mathcal{I}}$. Thus, any solution to the Dirac equation in the cavity can be spanned as:
\beq
\Psi(x,t) = \sum_{N=1}^{\infty}  \left( A_I \Psip_I(x,t)  +  B^*_I  \Psim_I(x,t) \right),\label{PsiDecomp}
\eeq
being $A_I = (\Psip_I | \Psi)$ and $B^*_I = (\Psim_I| \Psi)$.

The plus sign in \eqref{leftid} is quite remarkable because, as we will see, it encodes the Fermi-Dirac statistics of the system. Its presence is directly due to the positivity of the inner product in $\mathcal{S}$, \eqref{innerproduct}, a property that is not shared with the K-G inner product.

\subsection{Quantization}

Now, we can proceed to the quantization of the theory. To carry out the standard programme of canonical quantization, we promote the field $\Psi$ and its conjugate momentum $i\Psi^{\dagger}$ into operators satisfying the equal-time canonical \emph{anticommutation} algebra:
\beq
\lbrace \widehat{\Psi}_a(x,t), \widehat{\Psi}_b^{\dagger}(x',t)\rbrace = \delta(x-x') \delta_{a b}.\label{fieldalgebra}
\eeq
($a,\: b$ denote the component of the spinors we are considering).

The quantization process is completed finding a representation of the former algebra. Going back to the mode decomposition of the field, \eqref{PsiDecomp}, we arrive to:
\beq
{\Psi}(x,t) = \sum_I \left(  {A}_I \Psip_I(x,t) + {B}^{\dagger}_I \Psim_I(x,t) \right),
\eeq
where ${A}_I$ and ${B}^{\dagger}_I$ are now operators. The anticommutator \eqref{fieldalgebra} then translates into:
\beq
\lbrace  A_I, A_J^{\dagger} \rbrace = \delta_{I J}, \quad \lbrace  B_I, B_J^{\dagger} \rbrace = \delta_{I J}.\label{anticommrels}
\eeq
From that algebra it follows the customary creation-annihilation interpretation of the operators $A_I,\: A_I^{\dagger},\: B_I,\: B_I^{\dagger}$. Then, the Fock representation is constructed as follows:

First, the vacuum state $|0^G\ra$ is defined as the state annihilated by every annihilation operator:
\beq
 \ A_I |0^G\ra = 0, \ B_I |0^G\ra =0 \quad \forall I\in \mathds{N},\nn
\eeq
and a complete and orthonormal basis of field states is built by the states:
\begin{align}
|\lbrace N_I,\bar{N}_I\rbrace\ra= \prod_I (A_I^{\dagger})^{N_I} \prod_I (B_I^{\dagger})^{\bar{N}_I}  |0^G\ra,\label{Fockstates}
\end{align}
such as $\lbrace N_I,\bar{N}_I\rbrace$ is an infinite binary string informing of the occupation of the particle ($N_I$'s) and antiparticle ($\bar{N}_I$'s) modes. In order to be well defined states in the Fock space, they must satisfy the condition $ \sum_I N_I < \infty $. Furthermore, antisymmetrization is given by construction due to the CARs, \eqref{anticommrels}. Indeed, the states \eqref{Fockstates} are wedge products of one-particle states $|1_I\rangle = A_I^{\dagger} |0^G\rangle$ and $|\bar{1}_I\rangle = B_I^{\dagger} |0^G\rangle$, rather than tensor products. For example, a two-particle state is constructed as:
\begin{align}
|1_I, 1_J\rangle &= A_I^{\dagger} A_J^{\dagger} |0^G\rangle = |1_I\rangle \wedge |1_J\rangle \nn\\ &= \frac{1}{\sqrt{2}}\left( |1_I\rangle \otimes |1_J\rangle - |1_J\rangle \otimes |1_I\rangle \right).
\end{align}

The Hilbert space spanned by this basis, $\mathfrak{F}^G$, is nothing but the antisymmetrised Fock space associated with the one-particle Hilbert space $\mathfrak{H}:= Span[\lbrace |1_I\rangle,\: |\bar{1}_I\rangle \rbrace]$ \cite{Wald,Friis}:
\beq
\mathfrak{F}^G =   \bigoplus_{N=0}^{\infty} \bigotimes_{\quad a}^N \mathfrak{H}.
\eeq
The subindex $a$ refers to the fact that we are considering the antisymmetrized tensor product as we just have noted above.

 The evolution in this space is implemented by a self-adjoint Hamiltonian which can be expressed as:
 \beq
 H^G = \sum_I \Omega_I \left(A^{\dagger}_I A_I + B^{\dagger}_I B_I \right),
 \eeq
where the normal ordering prescription was considered  in order to regularize the infinite vacuum energy.

 Let us point out that the one particle Hilbert space was built by modes with definite momenta and therefore, completely delocalized in the cavity. Then, the resulting notion of a particle is related to a field excitation completely \emph{delocalized}. In the next section, a local particle notion will be developed.

\section{Manifestly local formalism}
\label{sec2}

The theory exposed up to this line gives us a global notion of particles with the associated creation-annihilation operators, and then, does not allow the existence of localized states within its formal body. Actually, it provides very few tools to deal with any notion of local operations. For example, what does this formalism tell us about what is happening in a portion of the cavity?

Imagine splitting the cavity $\mathcal{I}$ in two pieces, $[0,r]$ and $[r,R]$. Then, in complete analogy with \eqref{GM+}, \eqref{GM-}, in each subcavity we would leave with an orthonormal basis of stationary modes: $\lbrace \psip_i, \psim_i \rbrace$ in the left side and $\lbrace \tpsip_i, \tpsim_i \rbrace$ in  the right side.


Turning again into the study of the entire cavity, guided by the former stationary modes,  $\lbrace \psip_i, \psim_i \rbrace$, we define a set of (\emph{non-stationary}) modes solving the Cauchy problem defined by the following initial conditions:
\begin{widetext}\vspace{-4mm}
\beq
\psip_i (x, t=0) = \sqrt{\frac{\omega_i^2}{2 r (\omega_i^2 + \tfrac{m}{r})}}    \left(e^{i (p_i x + \delta_i)}\: u(p_i) - e^{-i (p_i x+\delta_i)}\: u(-p_i) \right) \Theta(r-x),\label{Cauchypr1}
\eeq
\beq
\psim_i (x, t=0) = \sqrt{\frac{\omega_i^2}{2 r (\omega_i^2 + \tfrac{m}{r})}}    \left(e^{-i (p_i x + \delta_i)}\: v(p_i) - e^{i (p_i x+\delta_i)}\: v(-p_i) \right) \Theta(r-x),\label{Cauchypr2}
\eeq\vspace{-5mm}
\end{widetext}
where now $p_i$ are the solutions to
\beq
\tan(p_i r) = -\frac{p_i}{m},\label{peq}
\eeq
and
\beq
\omega_i\equiv \sqrt{p_i^2 + m^2}, \qquad \delta_i\equiv \arctan{\left(\frac{p_i}{\omega_i + m}\right)}.
\eeq

There, $\Theta(x)$ is the Heaviside's step function.

The modes defined above are completely localized in $[0,\: r]$ at the initial time and then, they spread in the  cavity. Those modes at arbitrary time can be computed by means of the expansion in terms of the well known stationary modes $\lbrace \Psip_I, \Psim_I\rbrace$ constructed in the previous section:
\begin{align}
\psi^{(\pm)}_i (x, t) &= \sum_I   \left( (\Psip_I | \psi^{(\pm)}_i) \Psip_I(x,t) \right. \nn\\&\qquad\qquad \left. +  (\Psim_I | \psi^{(\pm)}_i) \Psim_I(x,t) \right) .
\end{align}
The explicit expressions for the coefficients $({\Psi}^{(\pm)}_I | \psi^{(\pm)}_i)$ are given  in appendix \ref{BogAp}, expressions \eqref{Ups++}, \eqref{Ups+-}. This solves the Cauchy problem posed by \eqref{Cauchypr1}, \eqref{Cauchypr2}.

\begin{figure}
\includegraphics[width=9cm]{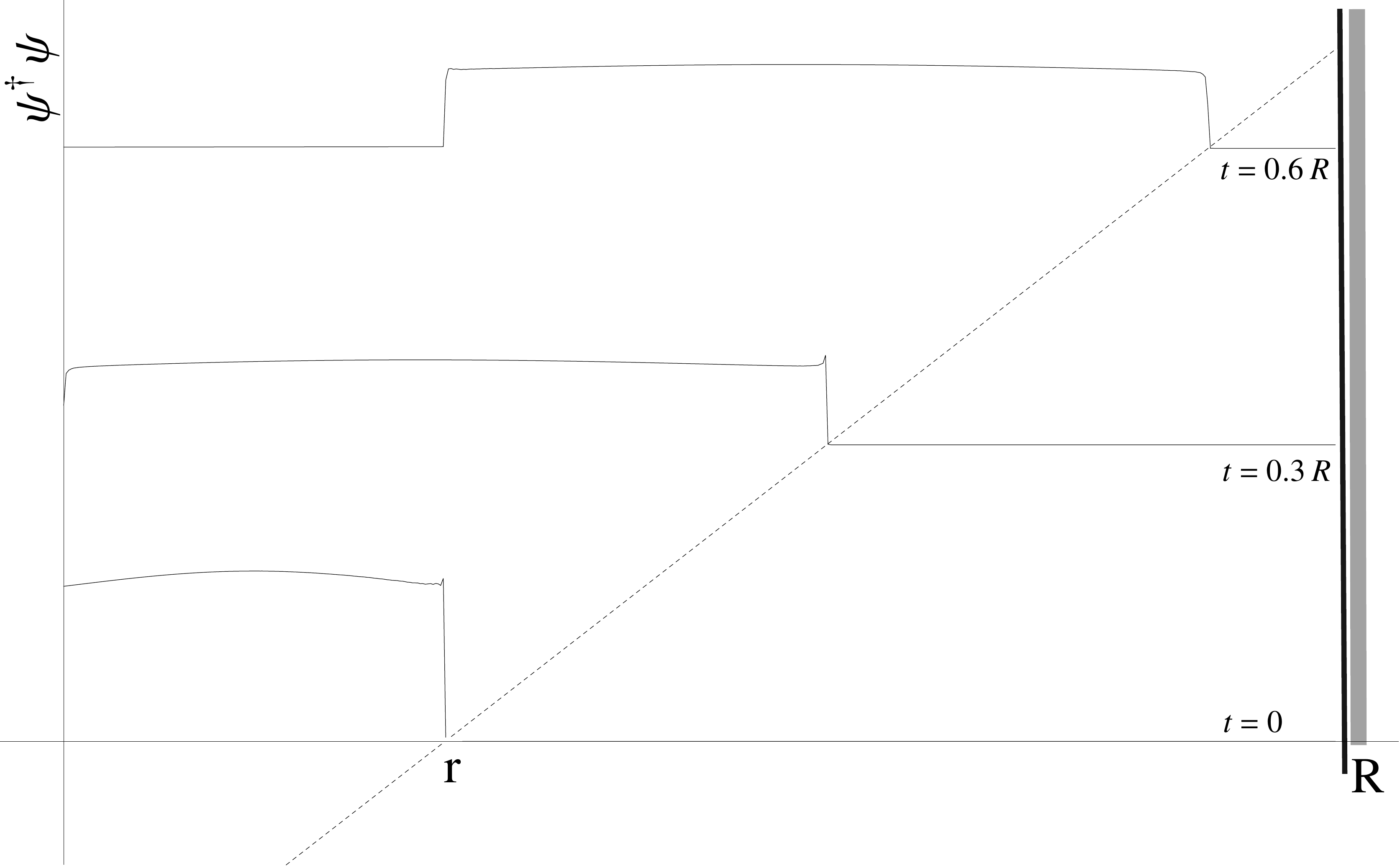}
\caption[a]{\small Causal evolution of local modes. Classical density ${\psi_1^{(\pm)}}^\dagger \psi_1^{(\pm)}$ in the case with $m=0.5 R,\: r=0.3 R$. The three plots correspond to three different times, $t=0,\: t=0.3 R,\: t=0.6 R$, while the dotted line is the light cone. The mixing of both positive and
negative frequencies allowed us to build up a localized mode spreading causally, avoiding the non-causal infinite tails that Hegerfeltd’s theorem would imply.}\label{fig1}
\end{figure}

 The numerical evaluation of the evolution of these modes is shown in figure \ref{fig1}. We see how the constructed modes spread causally inside the light cone. Let us point out that any other modes built from exclusively positive frequency solutions would spread instantaneously in the whole cavity, according to Hegerfeldt's theorem \cite{Hegerfeldt1998}, violating relativistic causality.

In complete analogy, similar modes are defined for the right side of the cavity, $\lbrace {\tpsip_i}, {\tpsim_i} \rbrace$.

\begin{figure}
\includegraphics[width=9cm]{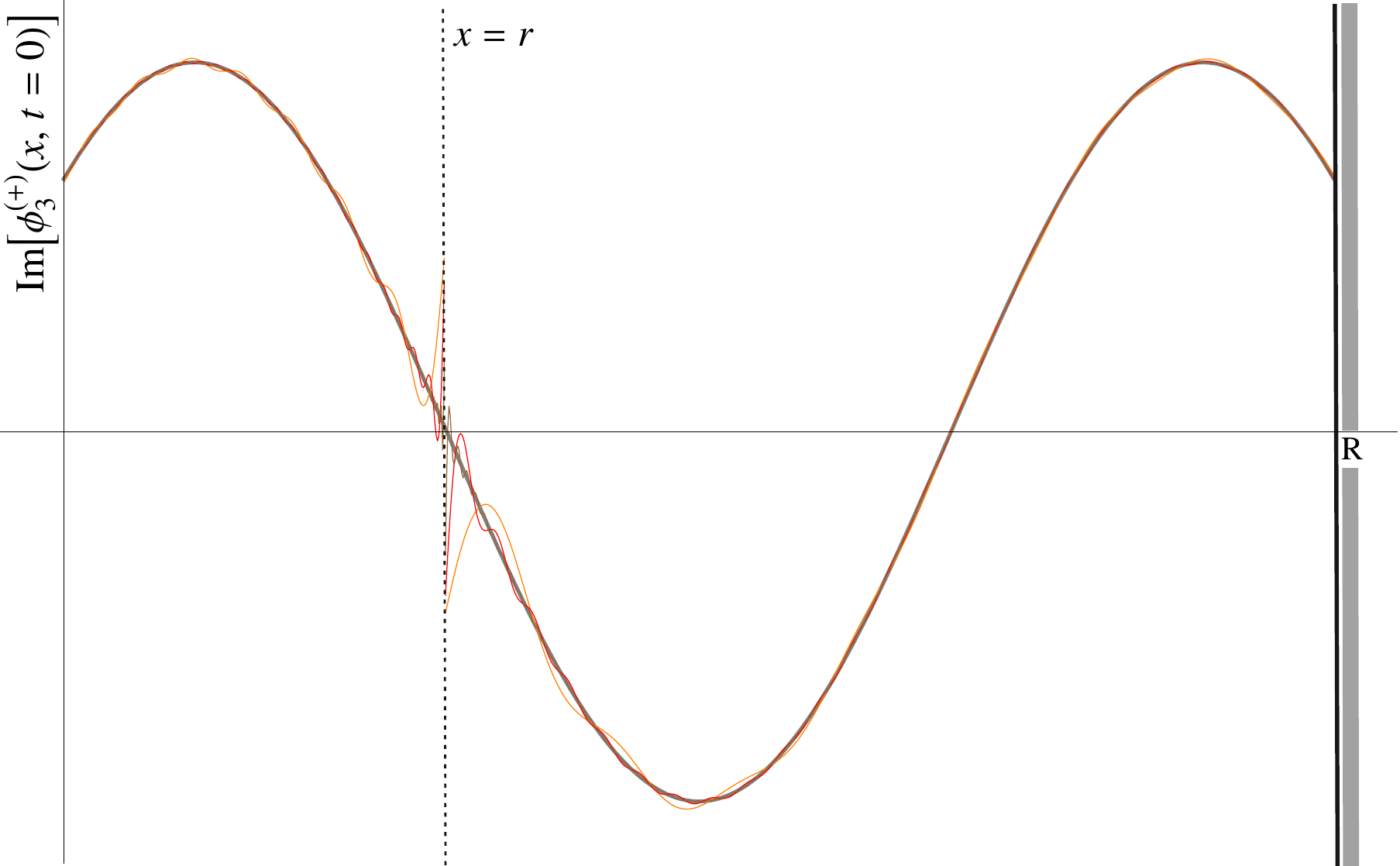}
\caption[a]{\small Expansion of the mode $\Psip_3$ in the local basis. Sum of the first fifteen (in orange), fifty (in red) and two hundred (in brown, hardly distinguishable from the actual curve, in gray) terms of the expansion for the first component.}\label{fig2}
\end{figure}

  Now, we notice that at the initial time we can construct in the subinterval $[0,r]$ any spinor subjected to the BCs \eqref{BCMIT} imposed at the boundaries $\partial \equiv \lbrace x=0,\ x=r \rbrace$. Furthermore, we can approximate any initial  condition  within this interval by an expansion in the modes $\lbrace \psip_i, \psim_i \rbrace$ with pointwise convergence everywhere except at the joining point $x=r$ (i.e., almost everywhere).   The same occurs for $\lbrace \tpsip_i, \tpsim_i \rbrace$ in $[r,\ R]$.  Considering both sets of modes $\lbrace \psip_i, \psim_i,\tpsip_i, \tpsim_i \rbrace$, we will be able to construct, in the same way, any initial condition in the entire interval $[0,R]$, again with pointwise convergence \emph{almost everywhere}. That is convergence in the norm induced by the inner product \eqref{innerproduct}.  So we can construct any initial condition in ${\mathcal{I}}$, and then, any solution to the Dirac equation in the cavity $\mathcal{I}$, $\Psi \in \mathcal{S}_{\mathcal{I}}$. In conclusion,  $\lbrace \psip_i,\psim_i,\tpsip_i,\tpsim_i \rbrace$ is another orthonormal basis of $\mathcal{S}_{\mathcal{I}}$ with pointwise convergence almost everywhere. We have a taste of how this works in figure \ref{fig2}.

Therefore, we can also span any solution $\Psi\in \mathcal{S}_{\mathcal{I}}$ as:
\begin{align}
\Psi (x,t) &= \sum_i \left( (\psip_i|\Psi)   \psip_i(x,t) +  (\psim_i|\Psi)   \psim_i(x,t) \right. \nn\\&\qquad
                        \left. + (\tpsip_i|\Psi)   \tpsip_i (x,t) +  (\tpsim_i|\Psi)   \tpsim_i(x,t) \right),\label{localdecomp}
\end{align}
having now the following completeness relation:
\begin{align}
\mathds{1} &= \sum_i \left(|\psip_i)(\psip_i| + |\psim_i)(\psim_i| \right.\nn\\&\left. \qquad +  |\tpsip_i)(\tpsip_i| +  |\tpsim_i)(\tpsim_i|\right).
\end{align}

\subsection{Local quantization}

From the construction of the solution space carried out above, follows another possible quantization of the system. Another representation of the field algebra \eqref{fieldalgebra} can be realized starting now from the decomposition in terms of local modes, \eqref{localdecomp}, yielding to:
 \begin{align}
 \Psi (x,t) &= \sum_i \left( a_i  \psip_i (x,t) + b^{\dagger}_i \psim_i(x,t)  \right. \nn\\
            &\qquad   \left. + a'_i \tpsip_i (x,t) + b'^{\dagger}_i \tpsim_i(x,t) \right),
 \end{align}
    providing now another  set of creation-annihilation operators $\lbrace a^{\dagger}_i, a_i, b^{\dagger}_i, b_i , a'^{\dagger}_i, a'_i, b'^{\dagger}_i, b'_i\rbrace$  satisfying the anticommutation relations:
\begin{align}
&\lbrace a_i, a^{\dagger}_j \rbrace =\delta_{i j}=  \lbrace b_i, b^{\dagger}_j \rbrace =\delta_{i j},\nn\\ &\lbrace a'_i, a'^{\dagger}_j \rbrace =\delta_{i j}= \lbrace b'_i, b'^{\dagger}_j \rbrace =\delta_{i j}.
\end{align}
The vacuum state this time will be the state satisfying:
\beq
a_i |0^L\ra  = 0=  b_i |0^L\ra,\ a'_i |0^L\ra  = 0= b'_i |0^L\ra \ \forall i\in \mathds{N}.
\eeq

The complete set of orthogonal states spanning the field states space, denoted now as $\mathfrak{F}^L$, is obtained applying the local creation operators in a similar way to \eqref{Fockstates} but  on the \emph{local} vacuum $|0^L\ra$.

As before, this space $\mathfrak{F}^L$ is the antisymmetrised Fock space
\beq
\mathfrak{F}^L =   \bigoplus_{n=0}^{\infty} \bigotimes_{\quad a}^n \mathfrak{H}^L ,
\eeq
but this time the one-particle  Hilbert space is identified with:
\begin{align}
\mathfrak{H}^L &= Span[\lbrace |1_i\rangle, |\bar{1}_i \rangle, |1'_i\rangle, |\bar{1}'_i \rangle \rbrace].
\end{align}

 Mimicking the standard procedure to regularize the vacuum energy, we can define a regularized Hamiltonian subtracting the contribution of the local vacuum:
 \beq
 H^L = H - \la 0^L | H | 0^L \ra,
 \eeq
where $H=\sum_I \Omega_I \left(A^{\dagger}_I A_I - B_I B^{\dagger}_I \right)$. At this point, this local quantization is formally completed.

 Notice that  the particle notion in this Fock space is associated to one of the both sides of the cavity, it is an actual \emph{local} quanta.

  \subsection{ Strictly localized states}

  Indeed, exploring the local one particle state:
\beq
|1_m\ra := a_m^{\dagger} |0^L\ra,\label{strictlylocalizedstate}
\eeq
we now can see that this is actually a  \emph{strictly localized} state. First of all, at the initial time we see that a local operator $\mathcal{Q}'$ acting on the region $[r,R]$ is build up by a series expansion of the operators $\lbrace a'_i,\: a'^{\dagger}_i,b'_i,\: b'^{\dagger}_i \rbrace$, that is, $\mathcal{Q}'= \mathcal{Q}'(a'_i, a'^{\dagger}_i,b'_i,\: b'^{\dagger}_i)$. Then, the anticommutation of $ a_i, \: a_j^{\dagger}$ with the set  $\lbrace a'_i,\: a'^{\dagger}_i,b'_i,\: b'^{\dagger}_i \rbrace$ guarantees the strict localization of $ |1_m\ra$ as we show below.

The operator $\mathcal{Q}'$ is build by products as:
 $$\mathcal{P}=\prod_s (a'_{s})^{i_s} (a'^{\dagger}_{s})^{j_s} (b'_{s})^{k_s} (b'^{\dagger}_{s})^{l_s}. $$
 For these products we realize that $$a_m \mathcal{P} = (-1)^{\#_{\mathcal{P}}} \mathcal{P} a_m$$ where $\#_{\mathcal{P}}= \sum_s (i_s+j_s+k_s+l_s)$. When calculating the average on the vacuum $|0^L\ra$, only terms with even powers   $\#_{\mathcal{P}}$ could give  non-vanishing contributions, and for that terms, $(-1)^{\#_{\mathcal{P}}}=1$. Then, we learn that:
\beq
\la 1_m | {\mathcal{Q}'}|1_m\ra  =  \la 0|  \mathcal{Q}'a_m a_m^{\dagger}|0\ra =\la 0| \mathcal{Q}'  |0\ra,\nn
\eeq
and $|1_m\ra$ is in fact a strict localized state in Knight's sense \cite{Knight}.

The causal spreading of the local modes depicted in figure \ref{fig1} also implies  that this state remains strictly localized in the light-cone $[0, r+ t]$.  In the Heisenberg picture, the explicit proof of this statement could be performed considering another family of local creators-annihilators  $\lbrace a''_i,\: a''^{\dagger}_i,b''_i,\: b''^{\dagger}_i \rbrace$  defined from non-stationary modes localized in another region $[r'',R]$ at a later time $t=\tau > 0$. Thus, these are local operators acting within the interval $[r'',R]$ at time $\tau$. Computing the anticommutators $\lbrace a_i, a''_i \rbrace,\ \lbrace a_i, a''^{\dagger}_i \rbrace$, etc.\footnote{ The computation is completely analogous to that in the case of scalar fields, changing commutators by anticommutators. See \cite{localquanta} for details.}, we realize that they vanish whenever $|r-r'' | > \tau$, i.e., whenever the region $[r'',R]$ at time $\tau$ is outside the light cone of the one particle state \eqref{strictlylocalizedstate}. Then, the argument followed above  also applies now to any local observable $\mathcal{Q}''_{\tau}$ defined over the region  $[r'',R]$ at time $\tau$. Accordingly, the state \eqref{strictlylocalizedstate} is strictly localized within the light cone $[0,r+t]$.

\section{Relating local and global descriptions}\label{bogCoefs}
\label{sec3}

Once this local formalism was successfully developed, it is specially interesting to explore the precise relation of our local representation of QFT with the standard representation constructed in section \ref{sec1}.

Spanning a general (classical) solution  $\Psi \in \mathcal{S}_\mathcal{I}$ in both orthonormal bases we have constructed in $\mathcal{S}_\mathcal{I}$:
\begin{align}
\Psi(x,t)&=\sum_I \left(A_I \Psip_I (x,t)+ B_I^* \Psim_I(x,t) \right) \nn\\
     &=    \sum_i \left( a_i \psip_i(x,t) + b_i^* \psim_i(x,t) \right.  \nn\\&\left. \qquad \quad +a_i' \tpsip_i(x,t)  {b_i'}^* \tpsim_i(x,t) \right),\label{BogSpanning}
\end{align}
we realize that the transformation in the \emph{classical} solutions space is simply a change of bases. Indeed, the coefficients in both bases are related by a Bogoliubov transformation\footnote{More formally, a Bogoliubov transformation is a transformation which preserves the classical symplectic structure, which is translated in the preservation of the canonical anticommutation (commutation for bosonic fields) relations.} which, taking into account the expressions  \eqref{Ups++}--\eqref{TUps+-}\footnote{See the appendix \ref{BogAp}}, can be cast consistently in the familiar form:
\begin{align}
 a_i &= \sum_I  \left(  \alpha^*_{i, I}  A_I - \beta^*_{i, I}    B^{*}_I \right),\label{bogrell1}\\
 b^{*}_i &= \sum_I \left( \alpha_{i, I}   B^{*}_I   - \beta_{i, I}  A_I \right),\\
 a'_i &= \sum_I   \left( \alpha'^*_{i, I}  A_I  - \beta'^*_{i, I}   B^{*}_I \right),\\
 b'^{*}_i &= \sum_I \left( \alpha'_{i, I}   B^{*}_I   - \beta'_{i, I}  A_I \right).\label{bogrell4}
\end{align}

The coefficients $\alpha$ and $\beta$ are the Bogoliubov coefficients, and they completely characterize the transformation.

The orthonormality of local and global modes implies that these coefficients have to satisfy the necessary conditions:
\begin{align}
\sum_{I} \left( \alpha_{i, I}  \alpha^*_{j, I} +   \beta_{i, I} \beta^*_{j, I} \right) &=\delta_{i j}, \label{Bogcond1}\\
\sum_{I} \left( \alpha_{i, I} \beta_{j, I} +   \beta_{i, I} \alpha_{j, I} \right) &=0, \label{Bogcond2}\\
\sum_{i} \left( \alpha_{i, I} \alpha^*_{i, J} +   \beta_{i, I} \beta^*_{i, J} \right) &=\delta_{I J}, \label{Bogcond3}\\
\sum_{i} \left( \alpha_{i, I} \beta_{i, J} +   \beta_{i, I} \alpha_{i, J} \right) &=0, \label{Bogcond4}
\end{align}
and the same for the primed coefficients.
In the scalar case, the analog to these expressions carries a minus sign, again, due to the non-positivity of the K-G inner product.

This transformation in the classical solution space is straightforwardly translated into the quantum domain,  where now $a_i$, $b^{\dagger}_i,\ldots$ are operators, acting in a different Fock space than $A_I$, $B_I^{\dagger}$. Being more specific, in the quantum theory, the Bogoliubov transformation determined by (\ref{bogrel1}--\ref{bogrel4}) is a map between two different representations of the field algebra:
 \beq
  \mathfrak{B}:\: \mathfrak{F}^L \longrightarrow \mathfrak{F}^G.\nn
 \eeq
Then, a state in $\mathfrak{F}^L $, $|\Psi\ra$, is mapped in $\mathfrak{F}^G$ as $\mathfrak{B} |\Psi\ra$. Meanwhile, linear operators in $\mathfrak{F}^L$, $\mathcal{Q}$,  are mapped in linear operators in $\mathfrak{F}^G$. In matrix language, $\mathcal{Q} \longrightarrow \mathfrak{B} \mathcal{Q} \mathfrak{B}^{-1}$.

\subsection{Unitary inequivalence}\label{sec:unitaryineq}

The first task one could wonder about at this point, is if two different quantizations describe the same physical system. Turning into the quantum theory, if the map  which relates both representations ($\mathfrak{B}: \mathfrak{F}^L \rightarrow \mathfrak{F}^G$)
is an unitary map, every observable will take the same value in both quantizations and then, they will predict the same physical consequences \cite{Birrel}. In that case it is said that both representations are \emph{unitary equivalent}.

A necessary condition for the unitarity of the Bogoliubov transformation between both Fock spaces is given by \cite{labonte1974}:
\beq
\sum_{i,I} \left( |\beta_{i,I}|^2 + |\beta'_{i,I}|^2 \right) <\infty.\label{betatrace}
\eeq
In other words, $\beta$, understood as a matrix operator, must be Hilbert-Schmidt. As we show below, it is not our case.

Inspecting \eqref{Ups+-} and \eqref{TUps+-} we can conclude that row and column series are both convergent, i.e,
\begin{subequations}
\beq
\sum_{i} \left(|\beta_{i,I}|^2 + |\beta'_{i,I}|^2 \right) <\infty, \label{partialseriesa}
\eeq
\beq
\sum_{I} \left(|\beta_{i,I}|^2 + |\beta'_{i,I}|^2 \right) <\infty .\label{partialseriesb}
\eeq
\end{subequations}

This fact can be seen analyzing the behaviour of $|\beta_{i,I}|^2$ in the limits $i \rightarrow \infty$ and $I\rightarrow\infty$. $$ |\beta_{i,I}|^2\sim i^{-2}\text{ and }\sim I^{-2}\nn$$
 in each case. Then, the convergence of series \eqref{partialseriesa}, \eqref{partialseriesb} follows through the Maclaurin–Cauchy convergence test. The situation now is different from what happens for a scalar field, where the respective sum over $i$ was always divergent \cite{localquanta}.  Nevertheless, it does not mean that \eqref{betatrace} is convergent. The convergence of the double serie \eqref{betatrace} requires more attention.

The Abel’s ($k,l$)-th Term Test states the following necessary condition for the convergence of a double serie $\sum_{l,k} a_{l,k}$:
\beq
\lim_{k,l \rightarrow\infty} k \:l \: a_{k,l} = 0.
\eeq

In the appendix \ref{LimAp} it is shown that the limit $\lim_{i,I \rightarrow\infty} i I |\beta_{i,I}|^2 \ne 0$. Indeed, this limit does not exist since his value depends on the way in which the limit is taken.

Finally, we are led to conclude that both, Fock quantization and that constructed here, are unitarily inequivalent. This vindicates the early Knight results \cite{Knight}, who showed that strictly localized states can not be achieved using a finite number of Fock quanta.

\section{Local properties of the global vacuum}
\label{sec4}

The relations \eqref{bogrell1}--\eqref{bogrell4} enable us to act on the ordinary Fock space $\mathfrak{F}^G$ with the local operators. We choose  the action of $a_i$ on an arbitrary state of the Fock basis as an illustrative example to check whether these operators are well defined  in $\mathfrak{F}^G$:
\begin{align}
\hspace{-2mm} a_i | \lbrace N_I, \bar{N}_I \rbrace\ra = &\sum_J \left(\alpha_{i, J}^* A_J | \lbrace N_I, \bar{N}_I \rbrace\ra
-\beta_{i, J}^* B_J^{\dagger} | \lbrace N_I, \bar{N}_I \rbrace\ra \right),\nn
\end{align}
where   $\lbrace N_I, \bar{N}_I \rbrace $ is a binary string labeling the occupation number of the global modes. Then,  the norm of this state is:
\begin{align}
 \la \lbrace N_I, \bar{N}_I \rbrace| a^{\dagger}_i a_i |\lbrace N_I, &\bar{N}_I \rbrace \ra   \nn\\& = \sum_J \left(|\alpha_{i, J} |^2 \delta_{N_J 1} +  |\beta_{i, J} |^2 \delta_{\bar{N}_J 0} \right) \nn\\& \le \sum_J\left( |\alpha_{i, J} |^2 +  |\beta_{i, J} |^2 \right) = 1.\label{norm_a_n}
\end{align}
Which shows that it is a well defined state (up to a finite normalization) in $\mathfrak{F}^G$. Since this  is valid for any state of the Fock basis, it means that in fact the operators $a_i$ are well defined also in $\mathfrak{F}^G$.

Now, we can  explicitly compute the norm of $a_i$ as an operator on $\mathfrak{F}^G$:
\beq
||a_i|| = \underset{\la \psi | \psi \ra \le 1}{sup} \la \psi| a^{\dagger}_i a_i |\psi  \ra =1.
\eeq
Indeed, constructing the sequence of states $\left \lbrace |\psi_s \ra \equiv |\lbrace N_i = \theta_{[s-i]}, \bar{N}_i =0 \rbrace\ra \right\rbrace$ with the first $s$ particle modes occupied\footnote{ Here, $\theta_{[i]}$ is the discretized form of the Heaviside's step function.}, the supremum of $\la \psi_s| a^{\dagger}_i a_i |\psi_s  \ra$ saturates the inequality \eqref{norm_a_n}.

The same proof can be carried out for the other local operators. In addition, using the CAR $\lbrace a_i, a_i^{\dagger}\rbrace = \mathds{1}$, it is straightforward to see that:
\beq
||a_i^{\dagger} a_i|| = 1 =||a_i||^2,
\eeq
and again, in complete analogy, the same applies to any other of our local operators.
  Then, we can conclude that the sets $\lbrace a_i, a^{\dagger}_i, b_i, b^{\dagger}_i, a'_i, a'^{\dagger}_i, b'_i, b'^{\dagger}_i  \rbrace $ generate a \emph{$C^*$-- algebra} of local operators in $\mathfrak{F}^G$.

\subsection{Spectrum of local excitations}

Now we can study the \emph{local} number operators $$n_i=a^{\dagger}_i a_i,\: \bar{n}_i= b^{\dagger}_i b_i,\: n'_i=a'^{\dagger}_i a'_i,\: \bar{n}'_i= b'^{\dagger}_i b'_i$$ in $\mathfrak{F}^G$, exploring the local particle content of the global vacuum $|0^G\ra$ in terms of those operators. The average number of  local particles and antiparticles associated to the left side of the cavity is
:
\beq
\la 0^G| n_i|0^G\ra = \sum_I | \beta_{i,I}|^2 =\la 0^G|\bar{n}_i|0^G\ra.\nn
\eeq

\begin{figure*}
\includegraphics[width=8.5cm]{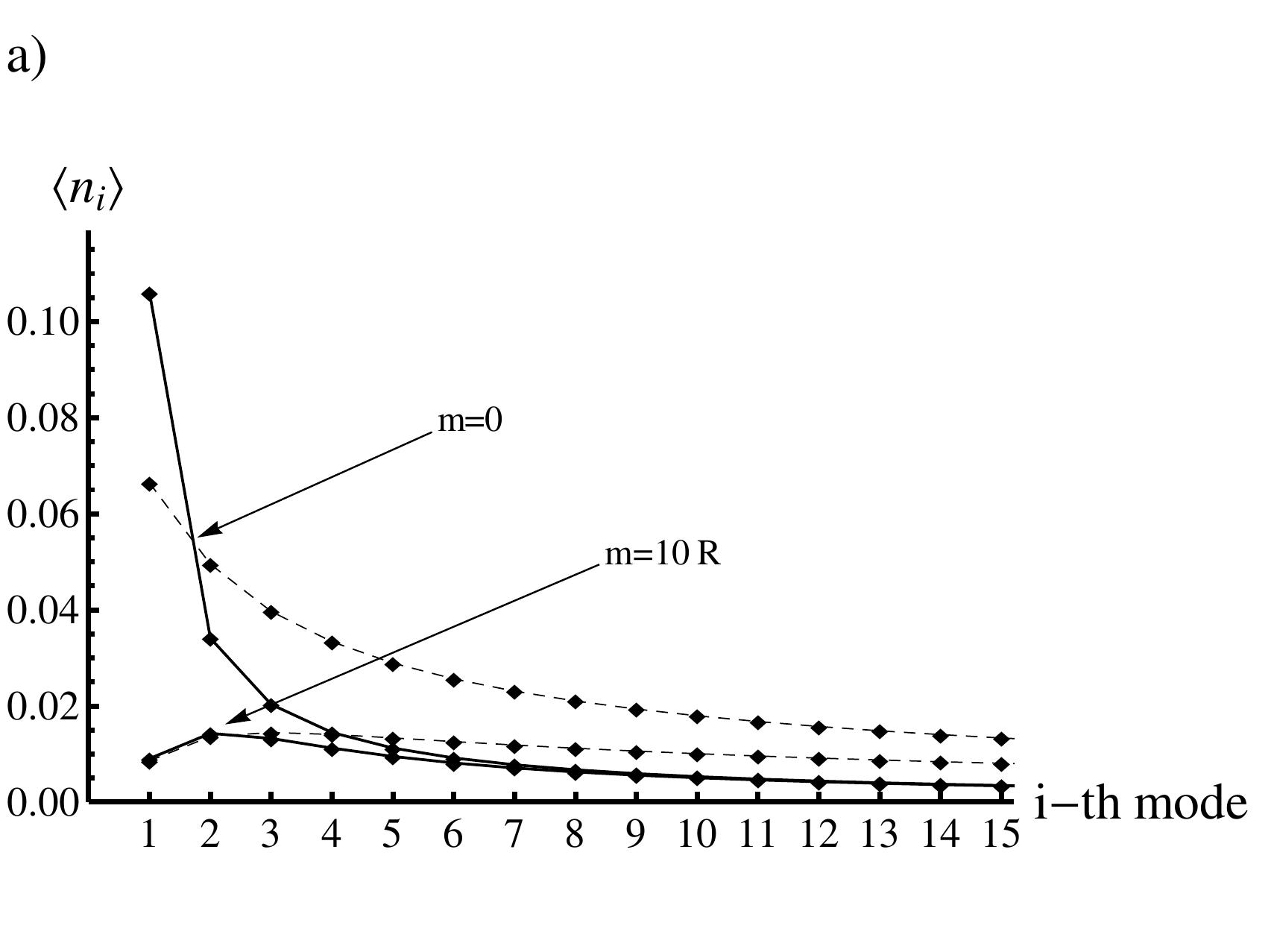}\qquad
\includegraphics[width=8.5cm]{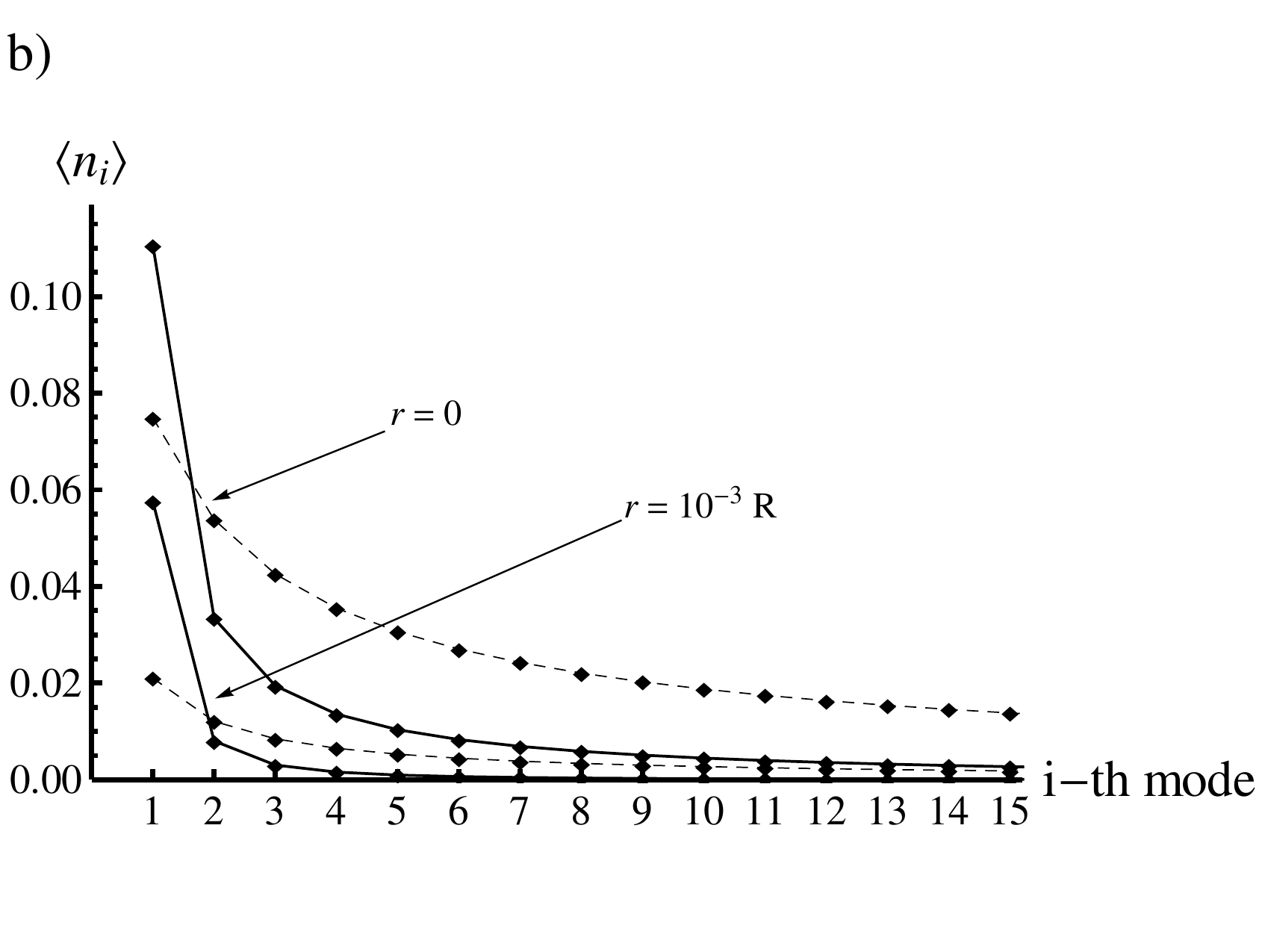}
\caption[a]{\small Local spectra in global vacuum $|0^G\ra$. Mean values $\la 0^G| n_i | 0^G\ra$ are represented for different masses of the field (m=0, m=10 R), with fixed $r=\frac{ R}{\pi}$(in the plot a)) and for different localization sizes, r (r=$5 \cdot 10^{-5}$ R, r=0.01 R, r=0.5 R), and fixed $m= R$ (in b)).}\label{fig3}
\end{figure*}

The coincidence between particle and antiparticle spectra was awaited due to the CPT invariance of the vacuum state, and it preserves the expected charge neutrality of this state.

 The intricate expression for local spectra can be evaluated numerically. Some examples are shown in figure \ref{fig3} in comparison with the case of a scalar field. The most significant difference  with the scalar case \cite{localquanta} has to do with the high frequency tails, which decrease faster in this case. Inspecting \eqref{Ups+-} we see that $|\beta_{i,I}|^2 \sim \omega_i^{-2}$ while in \cite{localquanta} $|\beta^{KG}_{i,I}|^2 \sim \omega_i^{-1}$. Other features of these spectra are quite similar to those in the scalar case resembling a tenuous thermal bath of local particles. This bath tends to disappear in the limits $r \rightarrow 0, R$ and $ m\rightarrow\infty$ since the coefficients $\beta_{i,I}$ vanish there.

 In a more general sense, we also can see from eq. \eqref{norm_a_n}  that the number of local excitations is bounded by $1$ for every state in $\mathfrak{F}^G$. Indeed, the equation \eqref{norm_a_n} is simply the expectation value of the observable $n_i$ in an arbitrary basis state. Then, this is also true for any normalised linear combination of these states. This fact is crucial for a consistent interpretation of these local observables as actual particle number operators.

\subsubsection{Trapping local quanta slamming down a mirror. Particle creation.}

As we have noted before, the local modes we have defined coincide with stationary modes when a mirror is placed  instantaneously at $x=r$. Then, if this action is actually implemented, the local modes become  actual stationary modes in the new cavities created and local excitations in the vacuum $|0^G\ra$ are revealed as real particles in the subcavities. This can be understood as a manifestation of the dynamical Casimir effect \cite{DynCasimir1,whatdoesitmeans}. As we have seen, due to unitary inequivalence, the total number of particles created is infinite, but this divergence is naturally regularized in practice by an ultraviolet cutoff related to the penetrability of the mirror and the finite velocity of its slamming \cite{BrownLouko2015}.

In any case, the message brought by our construction is that the local quanta created in the former gedanken-experiment are actually the local excitations existing in the vacuum $| 0^G\ra$. The mirror freezes this vacuum fluctuations revealing them as real particles.

\subsubsection{Particle creation by removing the mirror}
\label{removedmirror}

In section \ref{sec:unitaryineq} we have shown that:
\beq
\sum_{i} \left( |\beta_{i,I}|^2 + |\beta'_{i,I}|^2 \right),\nn
\eeq
is also convergent. This fact allows to study the operators $N_I= A^{\dagger}_I A_I,\ \bar{N}_I= B^{\dagger}_I B_I$ acting on $\mathfrak{F}^L$. Arguing similarly to the previous section, we can interpret the average values of
 these operators:
\begin{align}
\la 0^L | N_I |0^L\ra = \sum_i \left( |\beta_{i,I}|^2 + |\beta'_{i,I}|^2\right)  =\la 0^L | \bar{N}_I |0^L\ra  ,\nn
\end{align}
as the average number of particles (and antiparticles) created when a mirror placed at $x=r$ is suddenly removed. In the scalar case, this number was infinite for every frequency $\Omega_I$ \cite{localquanta} in abrupt contrast with the fermionic system discussed here. One can gain some insight on the interpretation of that phenomenon attending to the different statistics displayed by both fields. In the fermionic case treated here, it is not possible to create more than one particle with the same energy $\Omega_I$ and then, the number of such created particles is strictly limited. This fact is mathematically expressed in condition \eqref{Bogcond3}. This expression bounds the value of $\sum_i |\beta_{i,I}|^2<1$, constraining necessarily the asymptotic behaviour in the limit of large $i$, corresponding to the  limit large frequencies.

\subsection{Fluctuations and local excitations in the vacuum}

\begin{figure*}
\includegraphics[width=9cm]{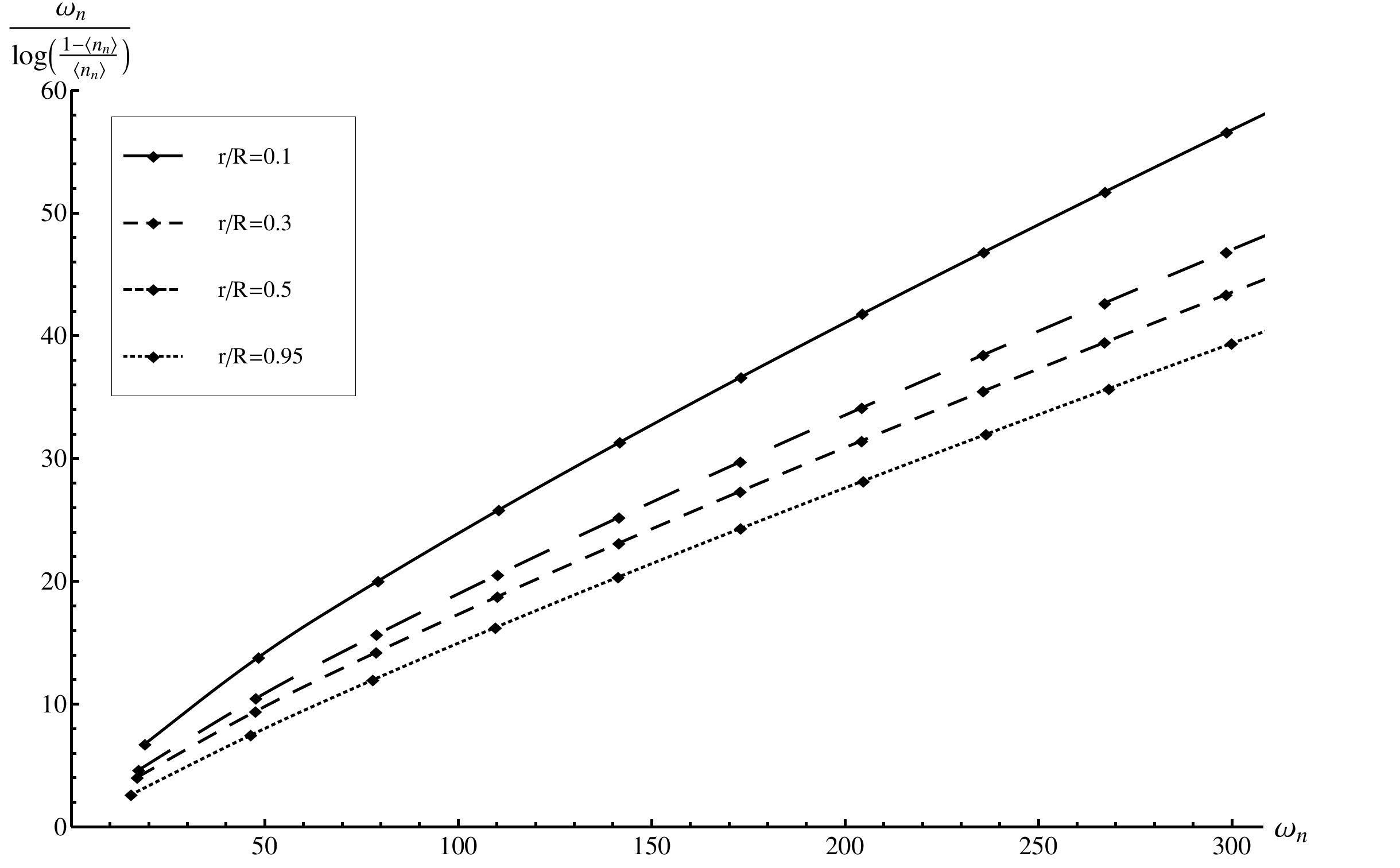}
\caption[a]{\small Plot of the transformed distribution ${ \omega_i}/{\log \left(\frac{1- \la n_i^{(a,b)}\ra}{\la n_i^{(a,b)}\ra}    \right)}$ clearly displaying a non Planckian behaviour.}\label{fig4}
\end{figure*}

It seems natural to interpret the local spectra $\la 0^G | n_i | 0^G \ra$, $\la 0^G | \bar{n}_i | 0^G \ra$ as a direct resemblance of vacuum fluctuations. We can gain some understanding of the nature of such fluctuations analysing more carefully the distribution they display. In particular, let us compare the distribution $\la n_i \ra$ with the thermal distribution for a fermionic system:
\beq
\la n (\omega) \ra =  \frac{1}{e^{\omega /  T} +1 },    \nn
\eeq
which is equivalent to the following equation for the temperature:
\beq
\frac{ \omega}{\log \left(\frac{1- \la n (\omega) \ra}{\la n(\omega) \ra}\right)} = T.
\eeq
Plotting the quotient ${ \omega_i}/{\log \left(\frac{1- \la n_i\ra}{\la n_i\ra}    \right)}$ against the frequency $\omega_i$ for the spectra  computed we can clearly see the non-thermal nature of the observed spectra. This is shown in figure \ref{fig4}. For a Planckian distribution we would expect an horizontal line in the plot, but the dependence with the frequency happens to be more complicate. In the limit $r=R$, $\la n_i\ra=0 $ and then ${ \omega_i}/{\log \left(\frac{1- \la n_i\ra}{\la n_i\ra}    \right)}=0$ as must be. In any other case, the resultant behaviour is the behaviour displayed in fig.\ref{fig4}.  The study of further properties of this spectrum remains for future work.

\subsection{Energy associated to local excitations}\label{SEC Energy}

What is a local excitation of the vacuum $|0^G\ra$? Such excitation can be realized as the action of our normalized local creators on $|0^G\ra$:
\begin{align}
|\varphi_i \ra &:= \frac{a^{\dagger}_i}{\sqrt{1 - \la 0^G| n_i|0^G \ra}} |0^G \ra \nn\\&= \frac{1}{\sqrt{1 - \la 0^G| n_i|0^G \ra}}\sum_I \alpha_{i,I} |1_I\ra.\label{quasilocalstate}
\end{align}

Now, we can compute the average energy of this state:
\beq
\la \varphi_i  | H^G | \varphi_i  \ra  =  \frac{1}{1 - \la 0^G| n_i|0^G \ra}  \sum_I \Omega_I |\alpha_{i,I}|^2=\infty.\label{quasilocalenergy}
\eeq
This comes from the fact that the behaviour of the term inside the sum behaves as $\sim \Omega_I^{-1} $ and then, using again the Maclaurin–Cauchy convergence test, it is divergent. To understand the origin of this divergence we have to turn our attention to the asymptotic behaviour of the coefficients $ \alpha_{i, I}, \beta_{i, I} $ in the high frequency limit. Exciting a local mode $\omega_i$ implies the excitation of the infinite number of global modes $\Omega_I$. The amplitude of this excitation is:
 $$
 \la 1_I | \varphi_i  \ra = \frac{ \alpha^*_{i,I} }{\sqrt{ 1 - \la 0^G| n_i|0^G \ra}}.
 $$
 This decays more slowly than in the scalar case, finally diverging. So, in physical grounds, the \emph{well defined} quasi-local state $| \varphi_i  \ra \in \mathfrak{F}^G$ is physically unattainable. Looking for some physical intuition, we can argue that to localize a fermion we have to struggle against the exclusion principle, making this task harder than for a bosonic system. This  intuition is dramatically confirmed by the divergent behaviour of the average value of the energy we have just seen.

  Nevertheless, in practice  a  quasi-local state can be constructed in a similar way than \eqref{quasilocalstate} but exciting only a finite number of global modes, many of them as we want. This state then will have a well defined finite energy, so the divergence \eqref{quasilocalenergy} does not seem really troubling.

  In complete analogy, it is also possible to consider the quasi-local states obtained acting with the other local operators in the vacuum:
\begin{align}
a_i |0^G \ra &= -\frac{1}{\sqrt{\la 0^G| n_i|0^G \ra}}\sum_I \beta^*_{i,I} |\bar{1}_I\ra,\nn \\
b^{\dagger}_i |0^G \ra &= \frac{1}{\sqrt{1-\la 0^G| \bar{n}_i|0^G \ra}}\sum_I \alpha_{i,I} |\bar{1}_I\ra,\nn \\
b_i |0^G \ra &=- \frac{1}{\sqrt{\la 0^G| \bar{n}_i|0^G \ra}}\sum_I \beta^*_{i,I} |1_I\ra.\nn
\end{align}
The analysis in these cases is similar to the one carried out just before, obtaining for all of them positive but divergent average energies. Notice that the annihilators $a_i,\ b_i$ produce also  excitations on the global vacuum $|0^G\ra$.

  What happens with the strictly localized states in $\mathfrak{F}^L$?
 The average energy computed above coincides (except for the constant factor $\frac{1}{1 - \la 0^G| n_i|0^G \ra}$) with the energy of a strictly local one-particle state in $\mathfrak{F}^L$,
\beq
\la 0^L |a_i H^L a^{\dagger}_i | 0^L \ra  =   \sum_I \Omega_I |\alpha_{i,I}|^2 = \infty.
\eeq
Then, the basis states of $\mathfrak{F}^L$ are unattainable from a physical point of view. We can put this pathological behaviour in analogy with the case of the basis states in the position representation of non-relativistic quantum mechanics. A Dirac delta state also displays a formal infinite average value for the energy, being these states unphysical in the same sense as our strictly localized states.


\subsection{Vacuum correlations and localizability}\label{SEC Corr}

The local operators constructed also can be used to study the spatial correlations displayed by the vacuum state.  The correlations between two observables $\mathcal{Q}$, $\mathcal{Q}'$ can be characterized by the normalized correlation function:
\beq
corr(\mathcal{Q}, \mathcal{Q}') = \frac{cov(\mathcal{Q}, \mathcal{Q}')}{\sqrt{cov(\mathcal{Q}, \mathcal{Q})}\sqrt{cov(\mathcal{Q}', \mathcal{Q}')}},\label{corr}
\eeq
where
\beq
cov(\mathcal{Q}, \mathcal{Q}') = \la 0^G | \mathcal{Q} \mathcal{Q}' | 0^G \ra - \la 0^G | \mathcal{Q}  | 0^G \ra \la 0^G | \mathcal{Q}' | 0^G \ra.
\eeq


We discriminate the correlations between  number of particles or antiparticles in both regions from the case in which we wonder about particles in one region but antiparticles in the other.

In the first case, expression \eqref{corr} is evaluated to:
\begin{align}
\hspace{-3mm}corr(n_l&, n'_k) =corr(\bar{n}_l, \bar{n}'_k) \nn \\ & = \frac{\sum_{I,J}  \alpha^*_{l, I} \alpha'_{k, I} \beta_{l, J} \beta'^*_{k J}}{\sqrt{\sum_{I,J}  |\alpha_{l, I}|^2 |\beta_{l, J}|^2 }\sqrt{\sum_{I,J}  |\alpha'_{k, I}|^2 |\beta'_{k, J}|^2 }  }. \label{corraa}
\end{align}

Correlations between particle number  in one side and antiparticle number in the other read as:
\begin{align}
\hspace{-3mm}corr(n_l&, \bar{n}'_k) = corr(\bar{n}_l, n'_k) \nn  \\ &= \frac{\sum_{I,J}  \alpha^*_{l, I} \beta'^*_{k, I} \beta_{l, J} \alpha'_{k J}}{\sqrt{\sum_{I,J}  |\alpha_{l, I}|^2 |\beta_{l, J}|^2 }\sqrt{\sum_{I,J}  |\alpha'_{k I}|^2 |\beta'_{k, J}|^2 }  }. \label{corrab}
\end{align}

\begin{figure*}
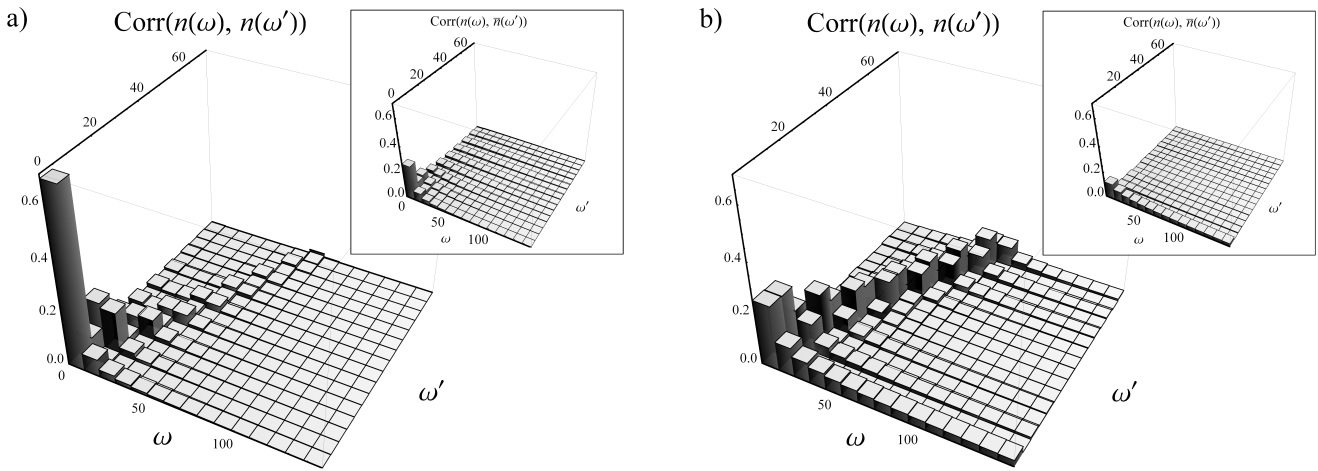

\begin{minipage}[l]{7.8cm}
\hspace{-2cm}\includegraphics[width=8.4cm]{figure6.pdf}
\end{minipage}
\begin{minipage}[r]{7.8cm}
\hspace{2cm}\includegraphics[width=8.4cm]{figure7.pdf}
\end{minipage}
\caption[a]{\small Absolute value of the correlations functions $corr(n_l,n'_k)$ displayed by the vacuum between space separated regions $[0,r]$ and $[r,R]$ where $\tfrac{r}{R}=\tfrac{1}{\pi}$ and the massless limit was taken. In the plot on the left we see the correlations for the Dirac field while on the right the case of a complex KG field is displayed. The main contribution to correlations comes from the correlation between particles states (or, equivalently, between antiparticle states) and they appear as the main content of the plots.  In the insets,  the  correlations between particles and antiparticle states or viceversa are  displayed.}\label{fig5}
\end{figure*}

These unwieldy expressions can be computed numerically.  In order to connect with the case of a scalar field we consider  a complex K-G field. In this case, it is easy to show that the correlation functions follow the expressions \eqref{corraa} and \eqref{corrab} but considering the Bogoliubov coefficients $\alpha_{i, I},\: \beta_{i, I}$ computed in \cite{localquanta}. For simplicity, we consider now the massless limit, which simplifies a lot the expressions we have to evaluate. Our numerical results for this limit are plotted in fig. \ref{fig5}. In both cases, scalars and fermions, most of the correlation \eqref{corraa} takes place between modes with similar frequencies. This contribution clearly dominates over \eqref{corrab}. In the fermionic case, the exclusion principle manifests again in the behaviour of large frequency modes. The correlations  decays more rapidly when the frequency increases, appearing the correlation between first modes enhanced.

%

How do all these relate to the question of localizability? Now, we can directly show that the quasi-local state built in this work fails to be strictly localized due to the nonvanishing spatial correlations displayed by global vacuum:
\begin{widetext}
\begin{align}
\la \varphi_l| n'_k | \varphi_l \ra -\la 0^G | n'_k | 0^G \ra &= \frac{\la 0^G|  a_l n'_k a^{\dagger}_l  | 0^G\ra }{1 - \la 0^G | n_l | 0^G \ra}   -    \la 0^G | n'_k | 0^G \ra \nn\\
    &=  \frac{   \la 0 ^G | n_l  |0^G \ra  \la 0 ^G |  n'_k |0^G\ra  - \la 0 ^G | n_l n'_k |0^G\ra }{ 1 - \la 0^G | n_l | 0^G \ra} =\frac{ - cov(n_l, n'_k) }{ 1 - \la 0^G | n_l | 0^G \ra} \propto corr(n_l, n'_k).
\end{align}
\end{widetext}
That is a direct proof of the non-strict localization of quasi-local states in $\mathfrak{F}^G$ but also it signals whom to blame of this fact: the spatial correlations pervading the vacuum.


\section{Conclusions}
\label{sec5}
In this work we constructed an alternative to the standard Fock quantization for free Dirac fields using creation-annihilation operators that –- instead of carrying a well defined momentum -- are localized in a  well defined region of space. Acting on the vacuum of this representation with these local creators gives rise to a natural notion of localized states that  turn out to be strictly localized within the region. We have
shown the unitary inequivalence between this representation and the standard Fock representation, a fact that reconciles the presence of our localized elementary excitations with the absence of strictly localized states in Fock space \cite{Knight,Licht1963,HalvorsonTh}. Up to this point, we have shown that fully strictly localized states can be constructed (actually, we have done so) but they do not lie within the state space of the standard Fock quantization. This idea, developed in \cite{localquanta} for scalar fields, is here adapted also for a fermionic  field.

The peculiarities of our quantization arise from the presence of local vacuum states, i.e., those states that vanish by the application of localized annihilation operators on them. A local vacuum state  $|0_V\rangle$ is associated to a region of space $V$ that does not contain any localized excitations, no matter what  happens outside $V$. However, this same region in the same state is occupied  with a swarm of Fock quanta. Local vacua in no way correspond to the Fock vacuum. The latter can be (and usually is) partitioned into sharp momentum vacua, but not into localized vacua. Thus, while the Fock vacuum state is customarily written as $|0^G\rangle = \bigotimes_k |0_k\rangle$, where k is momentum, it cannot be tensored as a product of local vacua  $|0^G\rangle \ne \bigotimes_V |0^L_V\rangle$ no matter how convoluted the choice of the set of regions $\lbrace{V}\rbrace$ is.

Focussing now on an arbitrary region $V$, we can construct the set of localized creation and annihilation operators and the corresponding local vacuum state. By using them, we obtain the Hilbert spaces $\mathcal{H}_V$ of strictly localized states and a local \emph{$C^*$} algebra $\mathcal{R}(V)$ associated to that  region. Notice that $\mathcal{H}_V$ is the invariant subspace of $\mathcal{R}(V)$. With the aid of the representations introduced  in this work we would have all the elements of an explicit GNS construction for the Hilbert space of the states that are strictly localized in $V$.  What prevents this program to succeed is unitary inequivalence; since $\sum_{i, I} |\beta_{i,I}| ^2 = \infty $ we can not normalize simultaneously $|0^L_V\rangle$ and the global vacuum of QFT $|0^G\rangle$ because they are related by an infinite factor. Notice also that this construction does not carry information whatsoever about what is going on outside V. In fact, the projectors that we may construct with the algebra $\mathcal{R}(V)$ are of infinite dimension when acting on the Hilbert space, $\mathfrak{F}^G$, of the QFT.

 What is the situation when applying our formalism to two spatially separated regions $V$ and $V'$?  There is no problem in duplicating the above construction for both regions. We may also build a cyclic vector $|0^L_{VV'}\rangle= |0^L_V\rangle \otimes |0^L_{V'} \rangle $ associated to our representation. Then, correlations involving both regions, i.e., of the type $$ \langle 0^L_{VV'}|\mathcal{Q}_V  \mathcal{Q}'_{V'} |0^L_{VV'} \rangle - \langle0^L_{VV'}|\mathcal{Q}_V  |0^L_{VV'}\rangle \langle0^L_{VV'}|  \mathcal{Q}'_{V'} |0^L_{VV'}\rangle \nn$$ being $\mathcal{Q}_V$, $\mathcal{Q}_{V'}$ operators in both separated regions,  are vanishing by construction.  Conversely, correlations in the global vacuum between both regions of space continue to be finite. As discussed in Sec. \ref{SEC Corr} this is to be blamed of the global character of the QFT vacuum.

A striking feature of the fermionic nature of the constituents of our local states is that they generically have infinite mean values for the energy. In Sec. \ref{SEC Energy} we have related this fact to the extra requirements needed for localization coming from the Pauli exclusion principle. In fact, in the case of scalar fields the  mere subtraction of the local vacuum energy allows to assign finite mean energy values to the local states of scalars \cite{localquanta}.
In Sec. \ref{sec4}, we have also shown that the action of local operators on the global Fock space is well defined. This was also true for scalar field but, differently than there, here we also find the same for the action of global operators on the local Fock space. This again can be seen as a consequence of the Pauli exclusion principle, as this comes from the plus sign on the left hand side of Eq. \eqref{Bogcond3} that was a minus sign in the case scalars (Eq. (15) of ref. \cite{whatdoesitmeans}).
\noindent This sign (coming from the positive definiteness of the inner product in the space of classical solutions) is the responsible for the fact that the mean value of the number operator for each local mode is bounded by 1 when acting on arbitrary global states. It uncovers a tight relation with the Pauli principle.

These operators allowed us to study some relevant properties of the global vacuum, e.g., the spectrum of local excitations and spatial correlations on this state. The imprint of the fermionic character of the system manifests in these properties when considering the high frequency limit, where we found that the mean values of number operators and correlations  are constrained as compared to the bosonic case. Conversely, low frequency modes result to be more strongly correlated.
\newpage

Summing up, we dealt here with the pervasive localization problem in QFT invoking the existence of unitarily inequivalent representations for free Dirac fields. Many issues remain still open. First of all, a detailed analysis of processes involving finite size detectors using our methods. Also, the generalization of the method introduced here to contexts involving more general metrics deserves to be explored. The long term goal would be the use of our local concept for particles where methods based on global spacetime symmetries are unsuitable.

 We cannot refrain from citing here the new proposals \cite{Single.Electron,Friis.Unlocking} showing (on a Gedanken experiment) that single electron states entangled between two spatially separated processes and, furthermore non local, may be observed with an appropriate setup. These states would correspond in our formalism to superpositions of the form
 $ \left(a^{\dagger}_V \otimes \openone_{V'} +  \openone_{V} \otimes a^{\dagger}_{V'} \right) | 0^G\rangle $, where the non-locality can be traced back to the collusion of local operators with the global vacuum. A main requirement for a sound analysis of single particle non-locality and/or entanglement \cite{Unruh.Evaporation,TanWalls91,VanEnk.Single} is that particle operators and modes participating in state superpositions have to belong or be associated to the relevant local algebras. We hope our treatment paves the way to explore this direction.
 
\section*{Acknowledgments}
This work was supported by Spanish MICINN Projects FIS2011-29287 and CAM research consortium QUITEMAD+ S2013/ICE-2801. A. M. Kubicki is supported by  the Spanish MINECO project MTM2014-57838-C2-2-P. H. Westman was supported by the CSIC JAE-DOC 2011 Program. The authors would like to thank Luis Garay for fruitful discussions about this work.

\clearpage

\appendix

\section{Dirac modes in the cavity}\label{modes}

In this appendix the stationary modes in the cavity $\mathcal{I}$, \eqref{GM+}, \eqref{GM-}, are computed.

Modes with definite positive frequency $\omega_p=\sqrt{p^2 + m^2}$ are constructed as the linear combination:
\beq
\Psip_p (x,t)= A^{(+)} e^{-i ( \omega t - p x)} u(p) + B^{(+)} e^{-i ( \omega t + p x)} u(-p),\nn
\eeq
while negative frequency modes are:
\beq
\Psim_p(x,t) = A^{(-)} e^{i ( \omega t - p x)} v(p) + B^{(-)} e^{i ( \omega t - p x)} v(-p),\nn
\eeq
being $A^{(\pm)},\ B^{(\pm)}$ some complex coefficients, determined by the BCs, and $u_{p},\ v_{p}$ are the spinors \eqref{eigenspinors}.

Explicitly, imposing the BCs \eqref{BCMIT} on these modes,  we obtain:
\begin{align}
\Psipm_p|_{x=0}&= i\gamma^1 \Psipm_p|_{x=0} \ \Rightarrow \ B^{(\pm)} = - A^{(\pm)} \frac{1\mp\frac{i p}{\omega + m}}{1\pm\frac{i p}{\omega + m}},\nn
\end{align}
and
\beq
\Psi^{(\pm)}_p|_{x=R}= -i\gamma^1 \Psi^{(\pm)}_p|_{x=R} \ \Rightarrow \ \tan(p R) = -\frac{p}{m}.\nn
\eeq

Therefore, modes in the cavity have the following structure:
\begin{align}
\Psip_I&(x,t)\nn\\
                    &  \hspace{-5mm} \propto (1+\frac{i P_I}{\Omega_I + m})U_{P_I}(x,t) - (1-\frac{i P_I}{\Omega_I + m})U_{-P_I}(x,t),\nn\\
\Psim_I&(x,t)\nn\\
                     &   \hspace{-5mm} \propto (1-\frac{i P_I}{\Omega_I + m})V_{P_I}(x,t) - (1+\frac{i P_I}{\Omega_I + m})V_{-P_I}(x,t) ,\nn
\end{align}
where  spectrum $\lbrace P_I \rbrace$ consist of the solutions to:
\beq
\tan(P_I R) = - \frac{P_I}{m}.\label{Peqq}
\eeq

Normalizing these modes with respect to the inner product \eqref{innerproduct} leads to \eqref{GM+}, \eqref{GM-}.

\section{Bogoliubov coefficients}\label{BogAp}

 Here we present the details and explicit expressions concerning the Bogoliubov transformation \eqref{bogrell1}--\eqref{bogrell4} between both quantizations considered in the paper.

It all starts in the spanning \eqref{BogSpanning}, which explicitly reads:
\begin{align}
\Psi(x,t)&=\sum_I \left((\Psip_I|\Psi) \Psip_I (x,t)+ (\Psim_I| \Psi) \Psim_I(x,t) \right) \nn\\
     &=    \sum_i \left((\psip_i|\Psi) \psip_i(x,t) + (\psim_i|\Psi) \psim_i(x,t) \right. \nn\\& \left.\quad+(\tpsip_i|\Psi) \tpsip_i(x,t) + (\tpsim_i|\Psi) \tpsim_i(x,t)\right).\nn
\end{align}\\

     Using the orthonormality of the set of local modes one obtains:
\begin{align}
(\psip_i|\Psi) &\equiv a_i = \sum_I   \left( (\psip_i|\Psip_I)  A_I + (\psip_i|\Psim_I)   B^{*}_I \right),\label{bogrel1}\\
(\psim_i|\Psi) &\equiv b^{*}_i = \sum_I \left( (\psim_i|\Psim_I)   B^{*}_I  +  (\psim_i|\Psip_I)  A_I \right),\label{bogrel2}\\
(\tpsip_i|\Psi) &\equiv a'_i = \sum_I  \left(  (\tpsip_i|\Psip_I)  A_I + (\tpsip_i|\Psim_I)   B^{*}_I \right),\label{bogrel3}\\
(\tpsim_i|\Psi) &\equiv b'^{*}_i = \sum_I \left( (\tpsim_i|\Psim_I)   B^{*}_I  +  (\tpsim_i|\Psip_I)  A_I \right) ,\label{bogrel4}
\end{align}
being $ A_I \equiv  (\Psip_I|\Psi), \ B^{*}_I \equiv (\Psim_I| \Psi) $ as in \eqref{PsiDecomp}.

$(\psi^{(\epsilon_1)}_i|\Psi^{(\epsilon_2)}_I),\ (\psi'^{(\epsilon_1)}_i|\Psi^{(\epsilon_2)}_I)$, where $\epsilon_i = +\text{ or }-$, are the Bogoliubov coefficients defining the transformation. The explicit computation of the inner products involved leads to the following expressions:
\begin{widetext}
\begin{align}
\alpha_{i, I} = ( \Psi^{(+)}_I | \psi^{(+)}_i ) &=  \phantom{ -i} \mathrm{C}_{i, I} \left[  \frac{p_i \Omega_I \sin P_I r  \cos p_i r - P_I \omega_i \cos P_I r \sin p_i r}{\Omega_I - \omega_i} + m \sin P_I r \sin p_i r   \right] =  ( \Psi^{(-)}_I | \psi^{(-)}_i),\label{Ups++}\\
\beta_{i, I} = - (  \Psi^{(-)}_I| \psi^{(+)}_i ) &=  - i \mathrm{C}_{i, I}\left[  \frac{p_i \Omega_I \sin P_I r  \cos p_i r + P_I \omega_i \cos P_I r \sin p_i r}{\Omega_I + \omega_i} + m \sin P_I r \sin p_i r  \right] = ( \Psi^{(+)}_I | \psi^{(-)}_i  ) ,\label{Ups+-}
\end{align}
and similarly, for the Bogoliubov coefficients related with the right partition of the cavity:

\begin{align}\vspace{-7mm}
\alpha'_{i, I}  &=  ( \Psi^{(+)}_I | \psi'^{(+)}_i  )\nn\\
                        &=  \phantom{- i} \mathrm{C}'_{i, I} \left[  \frac{p'_i \Omega_I \left(\sin P_I R  \cos p'_i (R-r)- \sin P_I r \right) - P_I \omega'_i \cos P_I R \sin p'_i (R-r)}{\Omega_I - \omega'_i} + m \sin P_I R \sin p'_i (R-r)   \right]\nn\\ &=  ( \Psi^{(-)}_I | \psi'^{(-)}_i  ) ,\label{TUps++}\\
\beta'_{i, I}  &=  - ( \Psi^{(-)}_I | \psi'^{(+)}_i ) \nn\\
                        &=   - i \mathrm{C}'_{i, I}\left[  \frac{p'_i \Omega_I \left( \sin P_I R  \cos p'_i (R-r) - \sin P_I r \right) + P_I \omega'_i \cos P_I R \sin p'_i (R-r)}{\Omega_I + \omega'_i} + m \sin P_I R \sin p'_i (R-r)  \right] \nn\\&= ( \Psi^{(+)}_I | \psi'^{(-)}_i )  ,\label{TUps+-}
\end{align}
\end{widetext}
where  were defined
\begin{align}
\mathrm{C}_{i, I}&= \sqrt{\frac{1}{rR (\omega_i^2 + m/r)(\Omega_I^2 + m/R) }},   \nn \\
\mathrm{C}'_{i, I}&= \sqrt{\frac{1}{(R-r) R (\omega_i^2 + m/(R-r))(\Omega_I^2 + m/R) }} . \nn
\end{align}

The properties of this inner products under the interchange of positive and negative frequency modes (interchange of the superscripts $+$ and $-$) and the definitions
\beq
\alpha_{i, I} = ( \Psi^{(+)}_I | \psi^{(+)}_i), \quad  \beta_{i, I} = - ( \Psi^{(-)}_I | \psi^{(+)}_i).
\eeq
makes the transformation \eqref{bogrel1}--\eqref{bogrel4} looks in the form \eqref{bogrell1}--\eqref{bogrell4}.

\section{Limits involved in  unitary inequivalence}\label{LimAp}

In this third appendix we show the explicit behaviour of $i I |\beta_{i,I}|^2$ when $i \text{ and } I$ tend to infinity.

First of all, observe that in such limit:

\begin{itemize}
\item $p_i, P_I >> m$, then, $\Omega_I \rightarrow P_I,\ \omega_i \rightarrow p_i$,
\item $\cos p_i r = - (\sin p_i r) m/p_i \:  \rightarrow 0$,
\item and,  even though, $n>>1 \Rightarrow p_i \simeq \frac{(2i-1)\pi}{2r}$,
\item  $\qquad\qquad\qquad\qquad N>>1 \Rightarrow P_I \simeq \frac{(2I-1)\pi}{2R}$.
\end{itemize}

Therefore, \eqref{Ups+-} behaves in that limit as:
\begin{align}
\beta_{i,I} &\sim   i \sqrt{\frac{1}{rR}}  \cos P_I r \sin p_i r \left[ \frac{ 1 }{P_I + p_i} + \frac{1 }{P_I  p_i}  \right]\nn\\
                    & \simeq     i \sqrt{\frac{1}{rR}}  \cos \left(\tfrac{(2 I-1) \pi r}{2 R}\right)(-1)^{i+1} \nn\\
                    &\qquad\qquad\qquad\qquad\qquad \times\left[ \frac{ r R }{(r I + R i )\pi} + \frac{r R }{I  i \pi}  \right].\nn
\end{align}

Thus,  the leading term of $i I |\beta_{i,I}|^2$ reads:
\begin{align}
&i I |\beta_{i,I}|^2\nn\\&\: \sim iI \: \frac{ r R}{\pi^2} \cos^2 \left(\tfrac{(2I-1) \pi r}{2 R}\right) \left[ \frac{ 1 }{ r I + R i } + \frac{ 1 }{ iI } \right]^2\nn\\
&\: \sim \frac{ r R}{\pi^2} \left[ \frac{ iI }{ r^2 I^2 + R^2 i^2 +2rR iI } + \frac{1 }{{iI}} + \frac{2iI}{r i I^2 + R i^2 I} \right].\nn
%
%
\end{align}

Inmediately we see that:
\beq
\lim_{i,I \rightarrow\infty} i I |\beta_{i,I}|^2 \ne 0,\label{limitUps-+}
\eeq
and actually, this limit does not exist. It is easy to see that the value of \eqref{limitUps-+} depends on how the limit is taken. Particularly, the inequality \eqref{limitUps-+} is clearly satisfied when the limit is taken along the path $i=I$. Otherwise, e.g.
\beq
\lim_{i,I=i^2 \rightarrow\infty} i I |\beta_{i,I}|^2 = 0.
\eeq

\bibliography{Bibliography}

\begin{thebibliography}{39}%
\makeatletter
\providecommand \@ifxundefined [1]{%
 \@ifx{#1\undefined}
}%
\providecommand \@ifnum [1]{%
 \ifnum #1\expandafter \@firstoftwo
 \else \expandafter \@secondoftwo
 \fi
}%
\providecommand \@ifx [1]{%
 \ifx #1\expandafter \@firstoftwo
 \else \expandafter \@secondoftwo
 \fi
}%
\providecommand \natexlab [1]{#1}%
\providecommand \enquote  [1]{``#1''}%
\providecommand \bibnamefont  [1]{#1}%
\providecommand \bibfnamefont [1]{#1}%
\providecommand \citenamefont [1]{#1}%
\providecommand \href@noop [0]{\@secondoftwo}%
\providecommand \href [0]{\begingroup \@sanitize@url \@href}%
\providecommand \@href[1]{\@@startlink{#1}\@@href}%
\providecommand \@@href[1]{\endgroup#1\@@endlink}%
\providecommand \@sanitize@url [0]{\catcode `\\12\catcode `\$12\catcode
  `\&12\catcode `\#12\catcode `\^12\catcode `\_12\catcode `\%12\relax}%
\providecommand \@@startlink[1]{}%
\providecommand \@@endlink[0]{}%
\providecommand \url  [0]{\begingroup\@sanitize@url \@url }%
\providecommand \@url [1]{\endgroup\@href {#1}{\urlprefix }}%
\providecommand \urlprefix  [0]{URL }%
\providecommand \Eprint [0]{\href }%
\providecommand \doibase [0]{http://dx.doi.org/}%
\providecommand \selectlanguage [0]{\@gobble}%
\providecommand \bibinfo  [0]{\@secondoftwo}%
\providecommand \bibfield  [0]{\@secondoftwo}%
\providecommand \translation [1]{[#1]}%
\providecommand \BibitemOpen [0]{}%
\providecommand \bibitemStop [0]{}%
\providecommand \bibitemNoStop [0]{.\EOS\space}%
\providecommand \EOS [0]{\spacefactor3000\relax}%
\providecommand \BibitemShut  [1]{\csname bibitem#1\endcsname}%
\let\auto@bib@innerbib\@empty
\bibitem [{\citenamefont {Knight}(1961)}]{Knight}%
  \BibitemOpen
  \bibfield  {author} {\bibinfo {author} {\bibfnamefont {J.~M.}\ \bibnamefont
  {Knight}},\ }\href@noop {} {\bibfield  {journal} {\bibinfo  {journal} {J.
  Math. Phys.}\ }\textbf {\bibinfo {volume} {2}},\ \bibinfo {pages} {459}
  (\bibinfo {year} {1961})}\BibitemShut {NoStop}%
\bibitem [{\citenamefont {Licht}(1963)}]{Licht1963}%
  \BibitemOpen
  \bibfield  {author} {\bibinfo {author} {\bibfnamefont {A.~L.}\ \bibnamefont
  {Licht}},\ }\href@noop {} {\bibfield  {journal} {\bibinfo  {journal} {J.
  Math. Phys.}\ }\textbf {\bibinfo {volume} {4}},\ \bibinfo {pages} {1443}
  (\bibinfo {year} {1963})}\BibitemShut {NoStop}%
\bibitem [{\citenamefont {Wallace}(2006)}]{Wallace2006}%
  \BibitemOpen
  \bibfield  {author} {\bibinfo {author} {\bibfnamefont {D.}~\bibnamefont
  {Wallace}},\ }\href {\doibase 10.1007/s11229-004-6248-9} {\bibfield
  {journal} {\bibinfo  {journal} {Synthese}\ }\textbf {\bibinfo {volume}
  {151}},\ \bibinfo {pages} {33} (\bibinfo {year} {2006})}\BibitemShut
  {NoStop}%
\bibitem [{\citenamefont {Deutsch}\ and\ \citenamefont
  {Hayden}(2000)}]{Deutsch2000}%
  \BibitemOpen
  \bibfield  {author} {\bibinfo {author} {\bibfnamefont {D.}~\bibnamefont
  {Deutsch}}\ and\ \bibinfo {author} {\bibfnamefont {P.}~\bibnamefont
  {Hayden}},\ }\href {\doibase 10.1098/rspa.2000.0585} {\bibfield  {journal}
  {\bibinfo  {journal} {Proceedings of the Royal Society of London A:
  Mathematical, Physical and Engineering Sciences}\ }\textbf {\bibinfo {volume}
  {456}},\ \bibinfo {pages} {1759} (\bibinfo {year} {2000})}\BibitemShut
  {NoStop}%
\bibitem [{\citenamefont {Swieca}(1965)}]{HaagSwieca1965}%
  \BibitemOpen
  \bibfield  {author} {\bibinfo {author} {\bibfnamefont {R.~H. J.~A.}\
  \bibnamefont {Swieca}},\ }\href@noop {} {\bibfield  {journal} {\bibinfo
  {journal} {Comm. Math. Phys.}\ }\textbf {\bibinfo {volume} {1}},\ \bibinfo
  {pages} {308} (\bibinfo {year} {1965})}\BibitemShut {NoStop}%
\bibitem [{\citenamefont {Schlieder}(1961)}]{ReedSchlieder1961}%
  \BibitemOpen
  \bibfield  {author} {\bibinfo {author} {\bibfnamefont {H.~R.~S.}\
  \bibnamefont {Schlieder}},\ }\href@noop {} {\bibfield  {journal} {\bibinfo
  {journal} {N. Cimento}\ }\textbf {\bibinfo {volume} {22}},\ \bibinfo {pages}
  {1051} (\bibinfo {year} {1961})}\BibitemShut {NoStop}%
\bibitem [{\citenamefont {Bialynicki-Birula}(1998)}]{BialynickiBirula1998}%
  \BibitemOpen
  \bibfield  {author} {\bibinfo {author} {\bibfnamefont {I.}~\bibnamefont
  {Bialynicki-Birula}},\ }\href {\doibase 10.1103/PhysRevLett.80.5247}
  {\bibfield  {journal} {\bibinfo  {journal} {Phys. Rev. Lett.}\ }\textbf
  {\bibinfo {volume} {80}},\ \bibinfo {pages} {5247} (\bibinfo {year}
  {1998})}\BibitemShut {NoStop}%
\bibitem [{\citenamefont {Terno}(2014)}]{Terno2014}%
  \BibitemOpen
  \bibfield  {author} {\bibinfo {author} {\bibfnamefont {D.~R.}\ \bibnamefont
  {Terno}},\ }\href {\doibase 10.1103/PhysRevA.89.042111} {\bibfield  {journal}
  {\bibinfo  {journal} {Phys. Rev. A}\ }\textbf {\bibinfo {volume} {89}},\
  \bibinfo {pages} {042111} (\bibinfo {year} {2014})}\BibitemShut {NoStop}%
\bibitem [{\citenamefont {Fulling}(1973)}]{Fulling1973}%
  \BibitemOpen
  \bibfield  {author} {\bibinfo {author} {\bibfnamefont {S.~A.}\ \bibnamefont
  {Fulling}},\ }\href {\doibase 10.1103/PhysRevD.7.2850} {\bibfield  {journal}
  {\bibinfo  {journal} {Phys. Rev. D}\ }\textbf {\bibinfo {volume} {7}},\
  \bibinfo {pages} {2850} (\bibinfo {year} {1973})}\BibitemShut {NoStop}%
\bibitem [{\citenamefont {Wald}(1994)}]{Wald}%
  \BibitemOpen
  \bibfield  {author} {\bibinfo {author} {\bibfnamefont {R.~M.}\ \bibnamefont
  {Wald}},\ }\href@noop {} {\emph {\bibinfo {title} {Quantum Field Theory in
  Curved Spacetime and Black Hole Thermodynamics (Chicago Lectures in
  Physics)}}}\ (\bibinfo  {publisher} {University of Chicago Press},\ \bibinfo
  {year} {1994})\BibitemShut {NoStop}%
\bibitem [{\citenamefont {Rovelli}(2004)}]{RovelliBook}%
  \BibitemOpen
  \bibfield  {author} {\bibinfo {author} {\bibfnamefont {C.}~\bibnamefont
  {Rovelli}},\ }\href@noop {} {\emph {\bibinfo {title} {Quantum Gravity}}}\
  (\bibinfo  {publisher} {Cambridge university Press},\ \bibinfo {address}
  {Cambridge},\ \bibinfo {year} {2004})\BibitemShut {NoStop}%
\bibitem [{\citenamefont {Theimann}(2008)}]{ThiemannBook}%
  \BibitemOpen
  \bibfield  {author} {\bibinfo {author} {\bibfnamefont {T.}~\bibnamefont
  {Theimann}},\ }\href@noop {} {\emph {\bibinfo {title} {Modern Canonical
  Quantum General Relativity}}}\ (\bibinfo  {publisher} {Cambridge university
  Press},\ \bibinfo {address} {Cambridge},\ \bibinfo {year} {2008})\BibitemShut
  {NoStop}%
\bibitem [{\citenamefont {Hegerfeldt}(1985)}]{Hegerfeldt1985}%
  \BibitemOpen
  \bibfield  {author} {\bibinfo {author} {\bibfnamefont {G.~C.}\ \bibnamefont
  {Hegerfeldt}},\ }\href@noop {} {\bibfield  {journal} {\bibinfo  {journal}
  {Phys. Rev. Lett.}\ }\textbf {\bibinfo {volume} {54}},\ \bibinfo {pages}
  {2395} (\bibinfo {year} {1985})}\BibitemShut {NoStop}%
\bibitem [{\citenamefont {Hegerfeldt}(1998)}]{Hegerfeldt1998}%
  \BibitemOpen
  \bibfield  {author} {\bibinfo {author} {\bibfnamefont {G.~C.}\ \bibnamefont
  {Hegerfeldt}},\ }\href {\doibase
  10.1002/(SICI)1521-3889(199812)7:7/8<716::AID-ANDP716>3.0.CO;2-T} {\bibfield
  {journal} {\bibinfo  {journal} {Annalen der Physik}\ }\textbf {\bibinfo
  {volume} {7}},\ \bibinfo {pages} {716} (\bibinfo {year} {1998})}\BibitemShut
  {NoStop}%
\bibitem [{\citenamefont {Malament}(1996)}]{Malament}%
  \BibitemOpen
  \bibfield  {author} {\bibinfo {author} {\bibfnamefont {D.}~\bibnamefont
  {Malament}},\ }in\ \href@noop {} {\emph {\bibinfo {booktitle} {Perspectives
  on Quantum Reality}}},\ \bibinfo {editor} {edited by\ \bibinfo {editor}
  {\bibfnamefont {R.}~\bibnamefont {Clifton}}}\ (\bibinfo  {publisher}
  {Kluwer},\ \bibinfo {year} {1996})\BibitemShut {NoStop}%
\bibitem [{\citenamefont {Halvorson}(2001)}]{HalvorsonTh}%
  \BibitemOpen
  \bibfield  {author} {\bibinfo {author} {\bibfnamefont {H.}~\bibnamefont
  {Halvorson}},\ }\href {http://philsci-archive.pitt.edu/346} {\emph {\bibinfo
  {title} {Locality, localization, and the particle concept: Topics in the
  foundations of quantum feld theory. PhD thesis, Pittsburg.}}}\ (\bibinfo
  {year} {2001})\BibitemShut {NoStop}%
\bibitem [{\citenamefont {Bialynicki-Birula}\ and\ \citenamefont
  {Bialynicka-Birula}(2009)}]{IwoSharp}%
  \BibitemOpen
  \bibfield  {author} {\bibinfo {author} {\bibfnamefont {I.}~\bibnamefont
  {Bialynicki-Birula}}\ and\ \bibinfo {author} {\bibfnamefont {Z.}~\bibnamefont
  {Bialynicka-Birula}},\ }\href {\doibase 10.1103/PhysRevA.79.032112}
  {\bibfield  {journal} {\bibinfo  {journal} {Phys. Rev. A}\ }\textbf {\bibinfo
  {volume} {79}},\ \bibinfo {pages} {032112} (\bibinfo {year}
  {2009})}\BibitemShut {NoStop}%
\bibitem [{\citenamefont {Vázquez}\ \emph {et~al.}(2014)\citenamefont
  {Vázquez}, \citenamefont {del Rey}, \citenamefont {Westman},\ and\
  \citenamefont {León}}]{localquanta}%
  \BibitemOpen
  \bibfield  {author} {\bibinfo {author} {\bibfnamefont {M.~R.}\ \bibnamefont
  {Vázquez}}, \bibinfo {author} {\bibfnamefont {M.}~\bibnamefont {del Rey}},
  \bibinfo {author} {\bibfnamefont {H.}~\bibnamefont {Westman}}, \ and\
  \bibinfo {author} {\bibfnamefont {J.}~\bibnamefont {León}},\ }\href
  {\doibase http://dx.doi.org/10.1016/j.aop.2014.07.031} {\bibfield  {journal}
  {\bibinfo  {journal} {Annals of Physics}\ }\textbf {\bibinfo {volume}
  {351}},\ \bibinfo {pages} {112 } (\bibinfo {year} {2014})}\BibitemShut
  {NoStop}%
\bibitem [{\citenamefont {Colosi}\ and\ \citenamefont
  {Rovelli}(2009)}]{ColosiRovelli2004}%
  \BibitemOpen
  \bibfield  {author} {\bibinfo {author} {\bibfnamefont {D.}~\bibnamefont
  {Colosi}}\ and\ \bibinfo {author} {\bibfnamefont {C.}~\bibnamefont
  {Rovelli}},\ }\href {\doibase 10.1088/0264-9381/26/2/025002} {\bibfield
  {journal} {\bibinfo  {journal} {Class. Quant. Grav.}\ }\textbf {\bibinfo
  {volume} {26}},\ \bibinfo {pages} {025002} (\bibinfo {year}
  {2009})}\BibitemShut {NoStop}%
\bibitem [{\citenamefont {Stone}(1930)}]{Stone1930}%
  \BibitemOpen
  \bibfield  {author} {\bibinfo {author} {\bibfnamefont {M.~H.}\ \bibnamefont
  {Stone}},\ }\href {http://www.ncbi.nlm.nih.gov/pmc/articles/PMC1075964/}
  {\bibfield  {journal} {\bibinfo  {journal} {Proc. Natl. Acad. Sci. U S A.}\
  }\textbf {\bibinfo {volume} {16}},\ \bibinfo {pages} {172} (\bibinfo {year}
  {1930})}\BibitemShut {NoStop}%
\bibitem [{\citenamefont {V.~Neumann}(1931)}]{v.Neumann1931}%
  \BibitemOpen
  \bibfield  {author} {\bibinfo {author} {\bibfnamefont {J.}~\bibnamefont
  {V.~Neumann}},\ }\href {\doibase 10.1007/BF01457956} {\bibfield  {journal}
  {\bibinfo  {journal} {Mathematische Annalen}\ }\textbf {\bibinfo {volume}
  {104}},\ \bibinfo {pages} {570} (\bibinfo {year} {1931})}\BibitemShut
  {NoStop}%
\bibitem [{\citenamefont {Brown}\ \emph {et~al.}(2015)\citenamefont {Brown},
  \citenamefont {del Rey}, \citenamefont {Westman}, \citenamefont {Le\'on},\
  and\ \citenamefont {Dragan}}]{whatdoesitmeans}%
  \BibitemOpen
  \bibfield  {author} {\bibinfo {author} {\bibfnamefont {E.~G.}\ \bibnamefont
  {Brown}}, \bibinfo {author} {\bibfnamefont {M.}~\bibnamefont {del Rey}},
  \bibinfo {author} {\bibfnamefont {H.}~\bibnamefont {Westman}}, \bibinfo
  {author} {\bibfnamefont {J.}~\bibnamefont {Le\'on}}, \ and\ \bibinfo {author}
  {\bibfnamefont {A.}~\bibnamefont {Dragan}},\ }\href {\doibase
  10.1103/PhysRevD.91.016005} {\bibfield  {journal} {\bibinfo  {journal} {Phys.
  Rev. D}\ }\textbf {\bibinfo {volume} {91}},\ \bibinfo {pages} {016005}
  (\bibinfo {year} {2015})}\BibitemShut {NoStop}%
\bibitem [{\citenamefont {Kialka}\ \emph {et~al.}(2016)\citenamefont {Kialka},
  \citenamefont {Ahmadi},\ and\ \citenamefont {Dragan}}]{Dragan2016}%
  \BibitemOpen
  \bibfield  {author} {\bibinfo {author} {\bibfnamefont {F.}~\bibnamefont
  {Kialka}}, \bibinfo {author} {\bibfnamefont {M.}~\bibnamefont {Ahmadi}}, \
  and\ \bibinfo {author} {\bibfnamefont {A.}~\bibnamefont {Dragan}},\ }\href
  {http://arxiv.org/pdf/1603.04890.pdf} {\bibfield  {journal} {\bibinfo
  {journal} {arXiv:1603.04890}\ } (\bibinfo {year} {2016})}\BibitemShut
  {NoStop}%
\bibitem [{\citenamefont {Chodos}\ \emph {et~al.}(1974)\citenamefont {Chodos},
  \citenamefont {Jaffe}, \citenamefont {Johnson},\ and\ \citenamefont
  {Thorn}}]{MITbag1}%
  \BibitemOpen
  \bibfield  {author} {\bibinfo {author} {\bibfnamefont {A.}~\bibnamefont
  {Chodos}}, \bibinfo {author} {\bibfnamefont {R.~L.}\ \bibnamefont {Jaffe}},
  \bibinfo {author} {\bibfnamefont {K.}~\bibnamefont {Johnson}}, \ and\
  \bibinfo {author} {\bibfnamefont {C.~B.}\ \bibnamefont {Thorn}},\ }\href
  {\doibase 10.1103/PhysRevD.10.2599} {\bibfield  {journal} {\bibinfo
  {journal} {Phys. Rev. D}\ }\textbf {\bibinfo {volume} {10}},\ \bibinfo
  {pages} {2599} (\bibinfo {year} {1974})}\BibitemShut {NoStop}%
\bibitem [{\citenamefont {Bordag}\ \emph {et~al.}(2001)\citenamefont {Bordag},
  \citenamefont {Mohideen},\ and\ \citenamefont {Mostepanenko}}]{Bordag20011}%
  \BibitemOpen
  \bibfield  {author} {\bibinfo {author} {\bibfnamefont {M.}~\bibnamefont
  {Bordag}}, \bibinfo {author} {\bibfnamefont {U.}~\bibnamefont {Mohideen}}, \
  and\ \bibinfo {author} {\bibfnamefont {V.}~\bibnamefont {Mostepanenko}},\
  }\href {\doibase http://dx.doi.org/10.1016/S0370-1573(01)00015-1} {\bibfield
  {journal} {\bibinfo  {journal} {Physics Reports}\ }\textbf {\bibinfo {volume}
  {353}},\ \bibinfo {pages} {1 } (\bibinfo {year} {2001})}\BibitemShut
  {NoStop}%
\bibitem [{\citenamefont {Friis}(2013)}]{Friis}%
  \BibitemOpen
  \bibfield  {author} {\bibinfo {author} {\bibfnamefont {N.}~\bibnamefont
  {Friis}},\ }\href@noop {} {\emph {\bibinfo {title} {Cavity Mode Entanglement
  in Relativistic Quantum Information. Phd. Thesis, Nottingham}}}\ (\bibinfo
  {year} {2013})\BibitemShut {NoStop}%
\bibitem [{\citenamefont {Gitman}\ and\ \citenamefont
  {Shelepin}(2001)}]{Gitman}%
  \BibitemOpen
  \bibfield  {author} {\bibinfo {author} {\bibfnamefont {D.~M.}\ \bibnamefont
  {Gitman}}\ and\ \bibinfo {author} {\bibfnamefont {A.~L.}\ \bibnamefont
  {Shelepin}},\ }\href {\doibase 10.1023/A:1004118431439} {\bibfield  {journal}
  {\bibinfo  {journal} {International Journal of Theoretical Physics}\ }\textbf
  {\bibinfo {volume} {40}},\ \bibinfo {pages} {603} (\bibinfo {year}
  {2001})}\BibitemShut {NoStop}%
\bibitem [{\citenamefont {Alonso}\ and\ \citenamefont {Vincenzo}(1997)}]{ven2}%
  \BibitemOpen
  \bibfield  {author} {\bibinfo {author} {\bibfnamefont {V.}~\bibnamefont
  {Alonso}}\ and\ \bibinfo {author} {\bibfnamefont {S.~D.}\ \bibnamefont
  {Vincenzo}},\ }\href {http://stacks.iop.org/0305-4470/30/i=24/a=018}
  {\bibfield  {journal} {\bibinfo  {journal} {Journal of Physics A:
  Mathematical and General}\ }\textbf {\bibinfo {volume} {30}},\ \bibinfo
  {pages} {8573} (\bibinfo {year} {1997})}\BibitemShut {NoStop}%
\bibitem [{\citenamefont {Stokes}\ and\ \citenamefont
  {Bennett}(2015{\natexlab{a}})}]{StokesBennett2}%
  \BibitemOpen
  \bibfield  {author} {\bibinfo {author} {\bibfnamefont {A.}~\bibnamefont
  {Stokes}}\ and\ \bibinfo {author} {\bibfnamefont {R.}~\bibnamefont
  {Bennett}},\ }\href {http://stacks.iop.org/1367-2630/17/i=7/a=073012}
  {\bibfield  {journal} {\bibinfo  {journal} {New Journal of Physics}\ }\textbf
  {\bibinfo {volume} {17}},\ \bibinfo {pages} {073012} (\bibinfo {year}
  {2015}{\natexlab{a}})}\BibitemShut {NoStop}%
\bibitem [{\citenamefont {Stokes}\ and\ \citenamefont
  {Bennett}(2015{\natexlab{b}})}]{StokesBennett1}%
  \BibitemOpen
  \bibfield  {author} {\bibinfo {author} {\bibfnamefont {A.}~\bibnamefont
  {Stokes}}\ and\ \bibinfo {author} {\bibfnamefont {R.}~\bibnamefont
  {Bennett}},\ }\href {\doibase http://dx.doi.org/10.1016/j.aop.2015.05.011}
  {\bibfield  {journal} {\bibinfo  {journal} {Annals of Physics}\ }\textbf
  {\bibinfo {volume} {360}},\ \bibinfo {pages} {246 } (\bibinfo {year}
  {2015}{\natexlab{b}})}\BibitemShut {NoStop}%
\bibitem [{\citenamefont {Birrel}\ and\ \citenamefont {Davies}(1982)}]{Birrel}%
  \BibitemOpen
  \bibfield  {author} {\bibinfo {author} {\bibfnamefont {N.~D.}\ \bibnamefont
  {Birrel}}\ and\ \bibinfo {author} {\bibfnamefont {P.~C.~W.}\ \bibnamefont
  {Davies}},\ }\href@noop {} {\emph {\bibinfo {title} {Quantum Fields in Curved
  Space}}}\ (\bibinfo  {publisher} {Cambridge university Press},\ \bibinfo
  {year} {1982})\BibitemShut {NoStop}%
\bibitem [{\citenamefont {Labonté}(1974)}]{labonte1974}%
  \BibitemOpen
  \bibfield  {author} {\bibinfo {author} {\bibfnamefont {G.}~\bibnamefont
  {Labonté}},\ }\href {http://projecteuclid.org/euclid.cmp/1103859661}
  {\bibfield  {journal} {\bibinfo  {journal} {Comm. Math. Phys.}\ }\textbf
  {\bibinfo {volume} {36}},\ \bibinfo {pages} {59} (\bibinfo {year}
  {1974})}\BibitemShut {NoStop}%
\bibitem [{\citenamefont {Moore}(1970)}]{DynCasimir1}%
  \BibitemOpen
  \bibfield  {author} {\bibinfo {author} {\bibfnamefont {G.~T.}\ \bibnamefont
  {Moore}},\ }\href {\doibase http://dx.doi.org/10.1063/1.1665432} {\bibfield
  {journal} {\bibinfo  {journal} {Journal of Mathematical Physics}\ }\textbf
  {\bibinfo {volume} {11}},\ \bibinfo {pages} {2679} (\bibinfo {year}
  {1970})}\BibitemShut {NoStop}%
\bibitem [{\citenamefont {Brown}\ and\ \citenamefont
  {Louko}(2015)}]{BrownLouko2015}%
  \BibitemOpen
  \bibfield  {author} {\bibinfo {author} {\bibfnamefont {E.~G.}\ \bibnamefont
  {Brown}}\ and\ \bibinfo {author} {\bibfnamefont {J.}~\bibnamefont {Louko}},\
  }\href {\doibase 10.1007/JHEP08(2015)061} {\bibfield  {journal} {\bibinfo
  {journal} {Journal of High Energy Physics}\ }\textbf {\bibinfo {volume}
  {8}},\ \bibinfo {pages} {1} (\bibinfo {year} {2015})}\BibitemShut {NoStop}%
\bibitem [{\citenamefont {Dasenbrook}\ \emph {et~al.}(2016)\citenamefont
  {Dasenbrook}, \citenamefont {Bowles}, \citenamefont {Brask}, \citenamefont
  {Hofer}, \citenamefont {Flindt},\ and\ \citenamefont
  {Brunner}}]{Single.Electron}%
  \BibitemOpen
  \bibfield  {author} {\bibinfo {author} {\bibfnamefont {D.}~\bibnamefont
  {Dasenbrook}}, \bibinfo {author} {\bibfnamefont {J.}~\bibnamefont {Bowles}},
  \bibinfo {author} {\bibfnamefont {J.~B.}\ \bibnamefont {Brask}}, \bibinfo
  {author} {\bibfnamefont {P.~P.}\ \bibnamefont {Hofer}}, \bibinfo {author}
  {\bibfnamefont {C.}~\bibnamefont {Flindt}}, \ and\ \bibinfo {author}
  {\bibfnamefont {N.}~\bibnamefont {Brunner}},\ }\href
  {http://stacks.iop.org/1367-2630/18/i=4/a=043036} {\bibfield  {journal}
  {\bibinfo  {journal} {New Journal of Physics}\ }\textbf {\bibinfo {volume}
  {18}},\ \bibinfo {pages} {043036} (\bibinfo {year} {2016})}\BibitemShut
  {NoStop}%
\bibitem [{\citenamefont {Friis}(2016)}]{Friis.Unlocking}%
  \BibitemOpen
  \bibfield  {author} {\bibinfo {author} {\bibfnamefont {N.}~\bibnamefont
  {Friis}},\ }\href {http://stacks.iop.org/1367-2630/18/i=6/a=061001}
  {\bibfield  {journal} {\bibinfo  {journal} {New Journal of Physics}\ }\textbf
  {\bibinfo {volume} {18}},\ \bibinfo {pages} {061001} (\bibinfo {year}
  {2016})}\BibitemShut {NoStop}%
\bibitem [{\citenamefont {Unruh}(1976)}]{Unruh.Evaporation}%
  \BibitemOpen
  \bibfield  {author} {\bibinfo {author} {\bibfnamefont {W.~G.}\ \bibnamefont
  {Unruh}},\ }\href {\doibase 10.1103/PhysRevD.14.870} {\bibfield  {journal}
  {\bibinfo  {journal} {Phys. Rev. D}\ }\textbf {\bibinfo {volume} {14}},\
  \bibinfo {pages} {870} (\bibinfo {year} {1976})}\BibitemShut {NoStop}%
\bibitem [{\citenamefont {Tan}\ \emph {et~al.}(1991)\citenamefont {Tan},
  \citenamefont {Walls},\ and\ \citenamefont {Collett}}]{TanWalls91}%
  \BibitemOpen
  \bibfield  {author} {\bibinfo {author} {\bibfnamefont {S.~M.}\ \bibnamefont
  {Tan}}, \bibinfo {author} {\bibfnamefont {D.~F.}\ \bibnamefont {Walls}}, \
  and\ \bibinfo {author} {\bibfnamefont {M.~J.}\ \bibnamefont {Collett}},\
  }\href {\doibase 10.1103/PhysRevLett.66.252} {\bibfield  {journal} {\bibinfo
  {journal} {Phys. Rev. Lett.}\ }\textbf {\bibinfo {volume} {66}},\ \bibinfo
  {pages} {252} (\bibinfo {year} {1991})}\BibitemShut {NoStop}%
\bibitem [{\citenamefont {van Enk}(2005)}]{VanEnk.Single}%
  \BibitemOpen
  \bibfield  {author} {\bibinfo {author} {\bibfnamefont {S.~J.}\ \bibnamefont
  {van Enk}},\ }\href {\doibase 10.1103/PhysRevA.72.064306} {\bibfield
  {journal} {\bibinfo  {journal} {Phys. Rev. A}\ }\textbf {\bibinfo {volume}
  {72}},\ \bibinfo {pages} {064306} (\bibinfo {year} {2005})}\BibitemShut
  {NoStop}%
\end{thebibliography}%

\end{document}